# Operando imaging of crystal structure and orientation in all components of all-solid-state-batteries

Quentin Jacquet[1#*], Jacopo Cele[2,3#*], Lara Casiez[2], Samuel Tardif[4], Asma Medjaheh[5], Stephanie Pouget[4], Manfred Burghammer[5], Sandrine Lyonnard[1*], Sami Oukassi[2*]

[1] Univ. Grenoble Alpes, CEA, CNRS, Grenoble INP, IRIG, SyMMES, 38000 Grenoble, France.

[2] Univ. Grenoble Alpes, CEA, Leti Grenoble F-38000, France

[3] ICMMO (UMR CNRS 8182) Univ. Paris-Sud Univ. Paris-Saclay Orsay 91190, France,

[4] University Grenoble Alpes, CEA, CNRS, IRIG, MEM, 38000 Grenoble, France.

[5] European Synchrotron Radiation Facility (ESRF), CS 40220, 71 Avenue des Martyrs, 38043 Grenoble, France

Corresponding authors: sandrine.lyonnard@cea.fr; sami.oukassi@cea.fr; quentin.jacquet@cea.fr; Jacopo.cele@cea.fr

# Equal contribution

**Abstract**

**A comprehensive understanding of interactions between cathode, electrolyte, anode, and packaging during battery operation is crucial for advancing performances but remains overlooked due to the lack of characterisation technics capable of measuring these components simultaneously. We perform a holistic investigation of a compact all-solid-state-battery using operando synchrotron X-ray micro-diffraction imaging. We image in real time and simultaneously the lattice parameter and crystal orientation of the dense $LiCoO_2$ cathode, the Ti current collector and the electrodeposited Li metal anode. We reveal that reaction mechanism of $LiCoO_2$ depends on the crystal orientation, and that, in dense electrodes as opposed to porous ones, the delithiation is limited by the formation of a Li-rich insulating interface. Li metal crystal orientation is found to be influenced initially by the Ti texture and to change within minutes during plating and stripping. These results demonstrate the power of X-ray imaging to link reaction mechanism and grain orientation during non-equilibrium processes.**

There is a strong drive to transition from Li-ion to solid-state batteries due to several advantages. Solid-state electrolytes are less flammable, minimizing thermal runaway risks, and have a higher $Li^+$ transference number, which reduces polarization and dendrite formation - key to enable high-energy-density Li metal anodes[1]. Energy density can be further enhanced by minimizing inactive components, such as conductive additives in electrodes, binders, current collectors, and the anode itself, as Li plating/stripping directly on the current collector offers the highest energy density[2–4]. The ultimate goal is an all-electrochemically active (AEA), anode-free solid-state battery[5,6]. Such design not only maximizes energy density but also minimizes electrode-electrolyte interfaces, reducing interphase formation that leads to polarization and degradation. Key challenges include (1) identifying factors that limit power and capacity in additive-free electrodes and (2) improving the reversibility of Li metal plating/stripping.

For the first question, slow diffusion is often cited as a limitation for charge transport across the electrode depth, causing Li concentration buildup at electrode-electrolyte or electrode-current collector interfaces[5,7]. Electron and $Li^+$ transport involve bulk and grain boundary contributions, along with tortuosity due to porosity and grain orientation. Identifying rate-limiting processes through electrochemical methods or modeling is possible but requires high-quality experimental validation[8], particularly operando characterization that maps Li concentration profiles and crystal orientation in real time. On the anode side, poor Li plating/stripping reversibility in solid-state batteries stems from issues like voids at the electrode/electrolyte interface, dendrites, or porous layers[9]. These morphologies depend on factors such as electrolyte composition, substrate nature, Li grain orientation, pressure, and current density[10]. However, limited techniques exist to characterize Li metal morphology, leaving electrodeposition mechanisms poorly understood and hindering targeted improvements[11]. Advancing this field requires real-time tracking of Li morphology and grain orientation across multiple crystals during plating/stripping.

There are many papers reviewing the need for more operando characterisations for solid-state batteries[12]. None of the reviewed technics seems to possess the perfect specifications to understand the limiting parameters of capacity, cycling stability and fast charging in AEA anode free batteries. Operando optical microscopy is not bulk sensitive and gives no direct information on the crystallographic orientation of the measured particles[13,14]. Operando electron microscopy requires small samples which might not be representatives[15–17]. X-ray tomography technics with time resolution compatible with operando characterisation are insensitive to grain orientation[9,18,19]. Globally, an operando technique capable of spatially resolving crystallographic grain structures and orientations within the components of a solid state battery while cycling is lacking.

In this work, we fill this gap and show how *operando* micro X-ray diffraction (µXRD) imaging at the synchrotron can provide holistic visualisation of dynamic processes limiting AEA anode free solid-state batteries. During a µXRD imaging experiment, 2D maps of the full battery are acquired in less than 5 minutes. Each pixel contains a 2D diffraction detector image giving access to crystalline structure and grain orientation. We study limiting process to fast charge and full discharge capacity in AEA $LiCoO_2$ electrode, archetype cathode material for portable electronics and micro devices. Moreover, we observe Li morphology and texture during 1st plating and stripping, while monitoring cell casing deformation. Beyond the important implication of these results for the design of higher energy density devices, this work opens the battery community to underutilized variety of fast scanning microbeam diffraction technics available at 4th generation synchrotrons, capable of measuring dynamic bulk properties of buried layers with unprecedented time and space resolutions in real batteries[20–23].

## Results

**µXRD imaging: a method capable of mapping phase and orientation distribution in multiple battery components**

Operando µXRD imaging is performed on a dense AEA anode-free solid-state microbattery (Fig. 1a) composed of a 20 µm dense LiCoO2 (LCO) cathode, 4 µm LIPON solid-state electrolyte, and 500 nm Ti acting as the anodic current collector on which Li plates during charge (Fig. 1b and 1c). The electrochemical stack is 40 µm thick x 2 mm wide x 350 µm long (z, x, y directions, respectively). It is scanned in front of a 1 x 1 µm X-ray beam, and diffracted X-rays are collected after the battery on a 2D detector. Maps of the full battery stack with 1 and 33 µm resolution along the z and x axis are measured in 5 min (Fig. 1d). Each pixel of the resulting map contains a detector image composed of Debye Scherrer rings which position on the detector informs on the lattice parameters of the crystalline materials scanned, hence allowing their identification. For example LCO, Ti or Au are mapped separately on Fig. 1e. Moreover, for LCO, Ti and Li, inhomogeneous intensity along the Debye Scherrer ring is observed which is characteristic of textured materials. For pristine LCO, exemplarily, Debye Scherrer ring corresponding to the (003) LCO Bragg reflection is composed of six intensity strips at different azimuths, corresponding to six LCO crystal orientations as depicted Fig. 1f. Four of the six orientations feature Li diffusion planes almost perpendicular to the electrode surface (LCO$^\perp$ - 20° between the electrode surface and the Li diffusion planes), ideal for Li diffusion out of the electrode. Some LCO crystals are oriented with the diffusion channel almost parallel to the film, labelled LCO$^\parallel$. Looking at the texture in the depth of the electrode (Fig. 1g), most of the film is composed of LCO$^\parallel$ and LCO$^\perp$ with approx. 1:1 ratio apart from the bottom, which features a significantly larger amount of LCO$^\perp$.

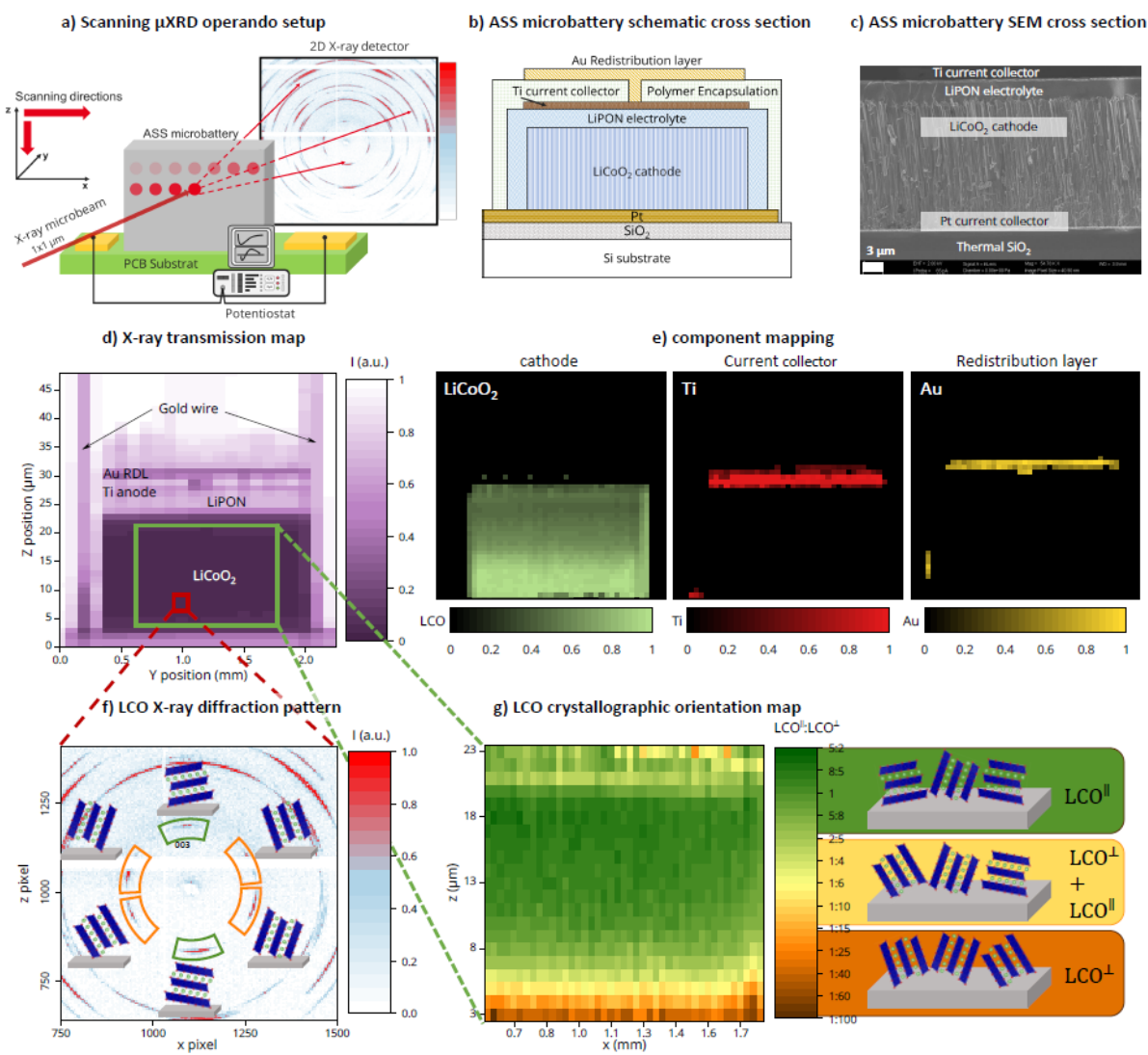

**Figure 1: Principle of operando µXRD imaging of all battery components.** a) Schematic of the operando µXRD imaging experiment. b) Schematic of the all solid state (ASS) battery. (c) Scanning electron microscopy image of the AEA cross section. (d) Typical X-ray transmission map obtained during the operando µXRD experiment in which all the different components can be identified based on their respective X-ray absorption. For example, LCO is the most absorbing element due to its thickness, density and relatively high atomic number. (e) Selective component (x,z) maps showing where the LCO, Ti and Au are located in the cell, obtained by selecting $(003)_{LCO}$, $(100)_{Ti}$, $(111)_{Au}$ diffraction rings, respectively, on the detector. For each pixel, the normalized intensity of each rings is represented using a color bar. (f) Typical 2D detector image, when the X-rays are focused on LCO, zoomed around the discontinuous $(003)_{LCO}$ Debye-Scherrer ring. It is composed of six intensity spots at different azimuths, namely: 25°, 160°, 205°, 345° (orange), 90° and 275° (green). Each intensity spot originates from a different crystal orientation as schematically represented on the panel. 'Green spots' correspond to $LCO^{\parallel}$ while 'orange spots' are $LCO^{\perp}$. (g) Map of the $LCO^{\parallel}:LCO^{\perp}$ peak intensity ratio in LCO electrode (left panel) together with a schematic of the LCO electrode texture as a function of depth (right panel).

**The effect of crystal orientation on cell parameter evolution in LiCoO$_2$ during delithiation**

The AEA battery is charged in 2.3 h using a constant voltage of 4.3 V (Fig. 2a). The current is almost constant during the first 45 minutes reaching up to 0.7 mAh.cm² (50% of the capacity). Final charging capacity reaches the expected value of 1.4 mAh.cm² (Fig. S1), corresponding to an average stoichiometry Li$_{0.5}$CoO$_2$. Cell parameter $c$ is determined from the (003)$_{LCO}$ position for a pixel during charge (Fig. 2b and Fig. S2). Maps of c parameter (Fig. 2c) evidence a clear biphasic transition at the beginning of charge for both LCO$^\perp$ and LCO$^\parallel$. Indeed, high and low c parameter phases coexist at the electrode level corresponding to the well-known insulator to metal transition between 0.95 < x < 0.75 for x in Li$_x$CoO$_2$[24]. However, c parameter changes from 14.0 Å to 14.13 Å for LCO$^\perp$ and 14.0 Å to 14.04 Å for LCO$^\parallel$ (Fig. 2d and 2e). In porous electrodes, LCO c parameter changes by 0.17±0.02 Å across the biphasic transition[24,25], which is consistent with the c parameter evolution of LCO$^\perp$ but not LCO$^\parallel$. During subsequent charging, c parameter of both LCO$^\perp$ and LCO$^\parallel$ increases homogeneously across the entire electrode, reaching at the end of charge 14.32 Å for LCO$^\perp$ and 14.12 Å for LCO$^\parallel$. To have a full picture of the lattice parameter variations, (110) diffraction spots for LCO$^\perp$ and LCO$^\parallel$ grains, observable on the detector at 90° and 210° azimuths, respectively, are analysed and $a$ lattice parameter variations determined (Fig. 2d and 2e, Fig. S3). LCO$^\perp$ and LCO$^\parallel$ have slightly different $a$ parameter values in the pristine state, namely 2.8035 Å and 2.803 Å respectively, but both decrease by 0.005 Å at the end of charge, which is comparable to literature[24,25]. Overall, $a$ parameter variation, which is directly connected to the CoO$_6$ shrinkage during delithiation and hence the state of charge, is similar for LCO$^\perp$ and LCO$^\parallel$, showing that both crystal orientations undergo similar delithiation level. That is consistent with the average composition calculated from electrochemistry. Therefore, the $c$ lattice parameter versus Li concentration relationship for LCO$^\parallel$ is substantially different from the LCO$^\perp$. The origin of the distinct phase diagrams for both crystal orientations remains unclear especially due to the lack of other *operando* characterisation experiments focusing on the effect of crystal orientation. We hypothesize that it could be due to mechanical effects (SI for discussion).

**What is limiting charge speed?**

Modelled and experimental Li concentration profiles across the depth of the electrodes during the biphasic reaction are compared Fig. 2f. Modelled Li concentration is obtained using a partial differential equation Newman-type model using the shape of the open circuit voltage curve and the diffusion coefficient as inputs[7]. Experimental Li concentration is estimated by calibrating c lattice parameter changes of LCO$^\perp$ with Li concentration using reference data from literature[24] (Fig. S4). The experiment confirms a series of modelling observation during the biphasic transition such as (1) position of the phase boundary, (2) its constant propagation speed of 10 nm.s$^{-1}$ and (3) concomitant out-of-equilibrium delithiation of the Li poor phase. Moreover, new insights are obtained from the experimental data. The same phase boundary speed is found for LCO$^\perp$ and LCO$^\parallel$. Additionally, the experimental interface between Li-rich and Li-poor phase is broader compared to the modelled one (Fig. 2f). The presence of a broad interface (6 μm) between Li-rich and Li-poor phases which cannot be predicted by modelling can be due to (1) particle by particle lithiation mechanism which has been observed for biphasic transition allowing to reduce interfacial energy[8,26], (2) low grain boundary diffusion and fast bulk diffusion, (3) heterogeneous tortuosity. After the phase transition, Li concentration is found homogeneous in LCO during the rest of delithiation, in agreement with modelling results. The absence of Li concentration gradients in the solid solution region shows that Li$^+$/e$^-$ diffusion is not limiting the speed of charge.

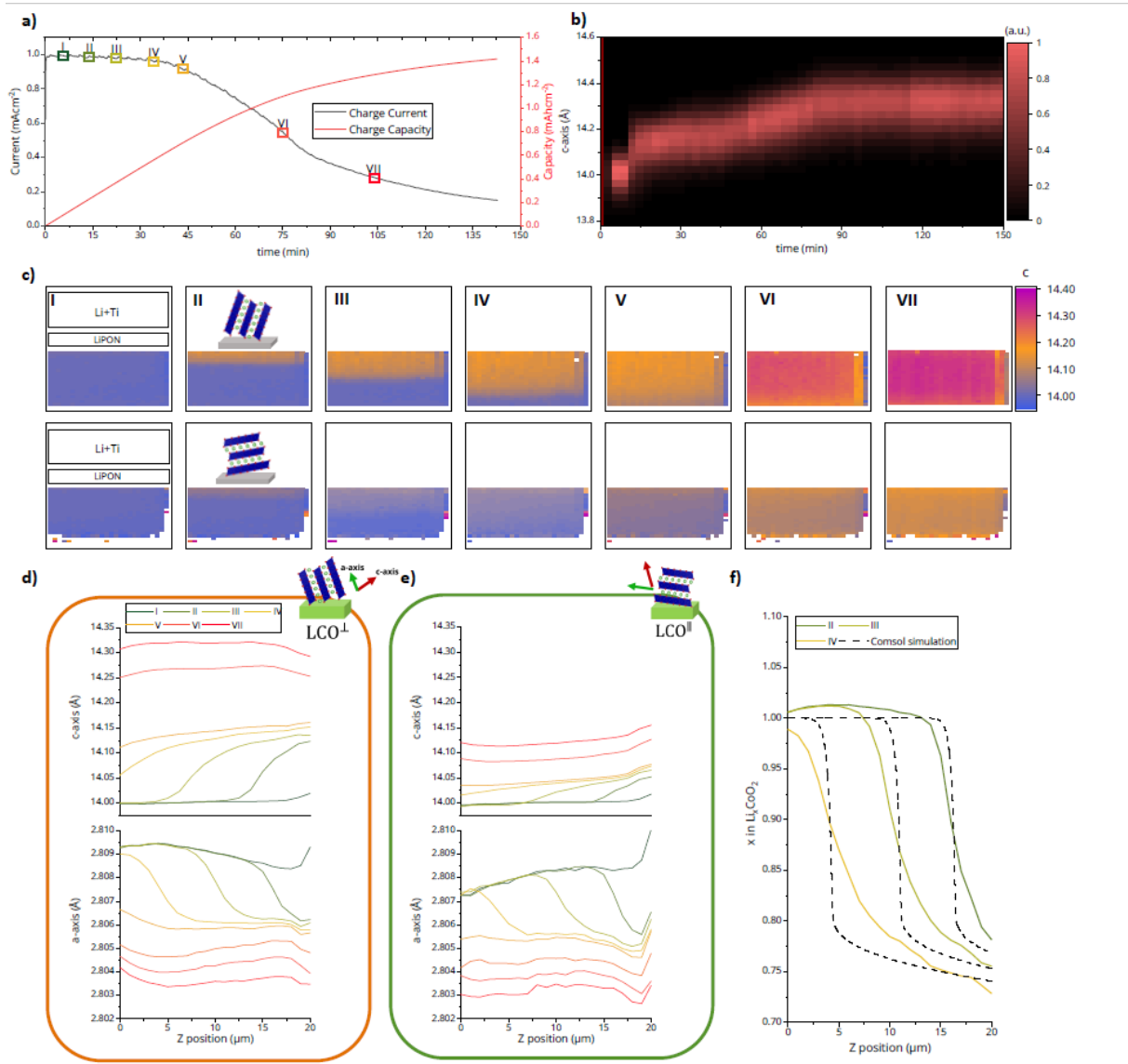

**Figure 2: LCO⊥ and LCO‖ structural mapping during charging.** a) Current (black) and capacity (red) of the AEA battery during potentiostatic charge at 4.3 V. b) Colour map of the *c* parameter as a function of time during charging (data is shown for a pixel at the top of the electrode). c) *c* parameter spatial maps measured for LCO⊥ and LCO‖ (top and bottom lines respectively) and shown for different times during the charge. Times, labelled as Latin numbers (I, II, III etc.), are shown in a). Each pixel is 33 x 1 µm in the horizontal and vertical direction, respectively. Map size is 1.36 mm x 40 µm (horizontal x vertical). d) and e) show the average *c* (top) and *a* (bottom) lattice parameter variations for LCO⊥ and LCO‖, respectively. Different coloured lines correspond to different time during the charge (I to VII). f) Experimental and modelled Li concentration profile during biphasic transition in plain and dotted line, respectively.

**What is limiting discharge capacity?**

Constant voltage charge is followed by a constant current discharge at i = 2 µA – corresponding to 4C (Fig. 3a). *c* parameters for both LCO⊥ and LCO‖ decrease constantly and homogeneously at the electrode level (Fig. 3b and 3c) showing the absence of Li concentration gradient during most of discharge. However, spatial heterogeneity is observed at the end of discharge, corresponding to the metal to insulator transition onset, with the insulating phase clearly observed at the top surface of the electrode (Fig. 3c) and appearing concomitantly with rapid decrease of voltage down to the cut-off voltage (3 V). During the rest following the discharge, the insulating phase progressively disappears. Phase fraction of insulating phase at the top pixels of the electrode reaches approx. 50% when the insulating phase is present over the first five surface pixels (5 microns). Modelled Li concentration and voltage curve also features a concomitant voltage drop and insulating phase formation towards the end of discharge, further suggesting that these two observations are correlated. Due to the low diffusion in the insulating phase, its surface rapidly reaches the maximum Li concentration shutting down charge transfer reaction. Li intercalation hence proceeds mainly through remaining Li-poor grains, which fraction reduces as the Li-rich phase nucleates. A smaller surface of Li-poor phase for a constant current leads to a drastic increase of the current density, and hence polarisation, driving the electrode potential down, ultimately reaching the cut-off voltage. The disappearance of the Li-rich phase during rest is surely due to Li concentration homogenisation in the film thickness, showing that Li-rich formation at this state of charge (SoC) and at 4C is kinetically driven as Li concentration reaches $Li_{0.75}CoO_2$. The formation of insulating phase at the electrolyte/electrode interface is limiting the discharge capacity of our AEA battery. Note that 1$^{st}$ cycle coulombic efficiency of other LCO AEAB is comparable to this work suggesting that it is a general mechanism in such device[27,28]. Interestingly, first cycle coulombic efficiency in LCO porous electrode Li-ion battery is much higher, and complete metal to insulator transition can be achieved in these systems even at very high C-rate[26,29,30]. This is explained by a higher Li-poor surface available for the charge transfer reaction due to higher interface area between active material and electrolyte in classical porous electrodes compared to all electrochemical active ones (Schematic Fig. 3g).

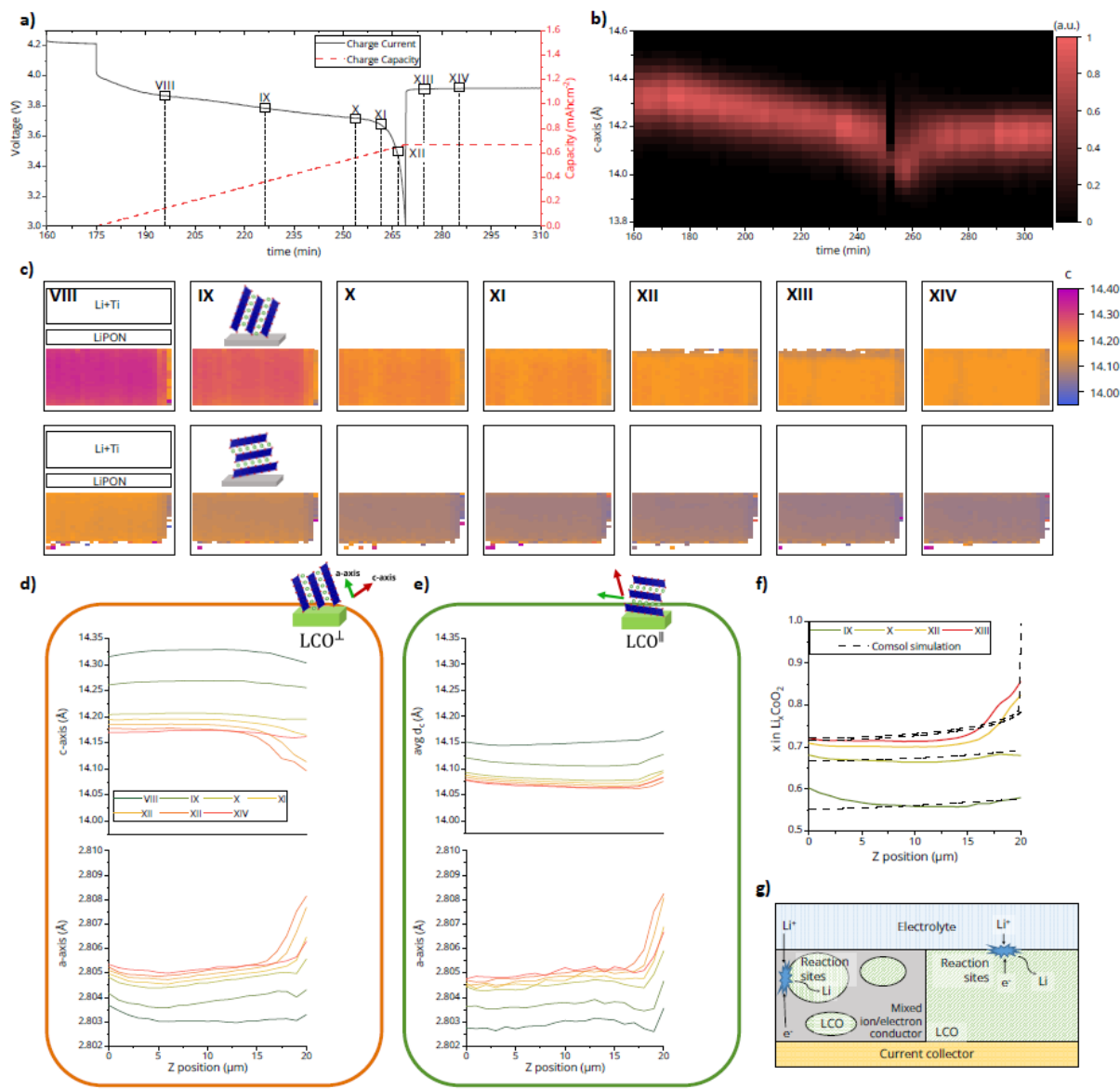

**Figure 3: LCO⊥ and LCO∥ structural mapping during discharging.** a) Voltage (black) and capacity (red) of the AEA battery during the galvanostatic discharge. b) Colour map of the averaged *c* parameter as a function of time during discharging (data is shown for a pixel at the top of the electrode). c) *c* parameter spatial maps measured for LCO⊥ and LCO∥ (top and bottom line respectively) and shown for different times during the discharge. Times, labelled as Latin numbers (VIII, IX, X etc.), are shown in a). Each pixel is 33 x 1 μm in the horizontal and vertical direction, respectively. Map size is 1.36 mm x 40 μm (horizontal x vertical). d) and e) show the average *c* (top) and *a* (bottom) lattice parameter variation for LCO⊥ and LCO∥, respectively. Different coloured lines correspond to different time during the discharge. f) Experimental and modelled Li concentration profile during biphasic transition in plain and dotted line, respectively. (g) Schematic showing the wider surface available for charge transfer in a porous electrode compared to an all-electrochemically active electrode[31]

**Imaging dynamic changes in Li metal texture and morphology**

At anode side, Ti current collector and Li metal diffraction signals are analysed in details starting with interpreting typical detector images. $(100)_{Ti}$ Debye-Scherrer rings feature an intense spot at 90° azimuth together with weaker intensity spots observed at 30°, 150°, 210° and 330° (Fig. 4a-left, Fig. S5). Hence, the chemically deposited Ti layer is textured with most of the $[100]_{Ti}$ perpendicular to the film surface (Fig. S6). Regarding Li metal, while $(110)_{Li}$, $(200)_{Li}$ and $(220)_{Li}$ are observable, analysis is focused on the highest intensity $(110)_{Li}$ (Fig. S7). A typical detector image after azimuthal integration (Fig. 4a-right) features narrow $(110)_{Li}$ spots showing the diffracted intensities originate from single crystals. Moreover, Li metal diffraction spots appear at a wide range of different azimuths without evident texture at first glance.

Next, Ti and Li are imaged in real space during plating (Fig. 4c). We observe that Ti current collector (red pixels) moves upward as Li plates underneath (green pixels) and its displacement exactly corresponds to the expected Li metal layer thickness (Fig. 4b). Interestingly at the end of charge, the Ti current collector appears thicker compared to the beginning of charge due to bending presumably induced by Li growth (Fig. S8). $(110)_{Li}$ intensity in the Li layer is not homogenous, featuring regions without any intensity (white regions in Fig 4c). This is not due to the absence of Li metal but the absence of grains in diffraction conditions. Moreover, the intensity is mostly present as vertical stripes frequently running all along the Li layer thickness. Finally, comparing the maps at 45 and 100 minutes (V and VII Fig 4c), corresponding to the middle and end of charge, respectively, it can be observed that grains are not at the same position, suggesting a complex dynamical evolution of the microstructure. These observations indicate that most of the Li metal grains are long, feature a unique crystallographic coherent domain all along the Li metal layer thickness and move/rotate during electrodeposition.

A more quantitative description is obtained by analysing each detector image (as a function of position and time), e.g. Li diffraction peaks are extracted, counted and their azimuths determined (see methods). Looking at the distribution of azimuths as a function of time (Fig. 4d), we find that there is a wide range of different azimuths, and hence crystal orientation, with some observable pattern suggesting an existing texture of the Li metal layer. Indeed, from 0 to 45 minutes most of the Li metal crystals have azimuths broadly gathered around 0°, 60°, 120° and 180°, with a clear absence of 30°, 90° and 150° oriented crystals. Beyond 45 minutes (0.80 mAh.cm²), the texture changes drastically since crystals with 30°, 90° or 150 ° azimuths are observed. These observations are consistent with Li being oriented initially with respect to Ti, following the epitaxial orientation described by Burgers orientation relationship $(001)_{Ti}//(110)_{Li}$ - $[100]_{Ti}//[111]_{Li}$ (Fig. 4g and SI)[32,33]. As the Li layer grows, the effect of Ti orientation is expected to weaken, in agreement with the observed change of texture after 0.8 mAh.cm² of Li plating (almost 4 microns of Li metal deposited). This result reveals the strength of *operando* µXRD texture imaging which enables tracking grain orientation and relate it with external parameters as substrate orientation or capacity (see SI for discussion).

Tracking the orientation of individual Li grains during plating/stripping is difficult because most grains rotate/move out of diffraction conditions in a few minutes ('ephemeral' grains in Fig. 4e). Some grains remain invariant over most of the plating ("stable" grains in Fig. 4e). To capture the motion of Li metal grains, total and new grains in diffraction conditions are counted (Fig. 4f). Total number of grains increases rapidly at the beginning of charge up to approx. 90 and decreases during the rest of the electrochemical sequence (80 grains - end of charge, 40 grains - discharge and rest). The number of new grain ranges from 30 – 10 in charge – discharge, respectively, hence representing a significant fraction of the observed grains ( > 30 %) showing the dynamic nature of the Li grains in the anode. Some of these observations can be rationalized in light of *ex situ* observation performed on liquid based batteries[34], since there is very little reports on the Li nucleation and growth in solid state batteries[35]. First, a high number of Li grains nucleate due to the relatively high current density at the beginning of the potentiostatic charge (1 mAh.cm²), then, as the Li metal layer grows and stack pressure increases, Li metal grains grow leading to the reduction of the grain number[36] (SI for more details). Being able to measure operando grains moving out of diffraction conditions calls for technical development to track grains as they move, hence having access to Li mechanical deformation, crucial to build a chemo-mechanical understanding of Li deposition in solid state batteries.

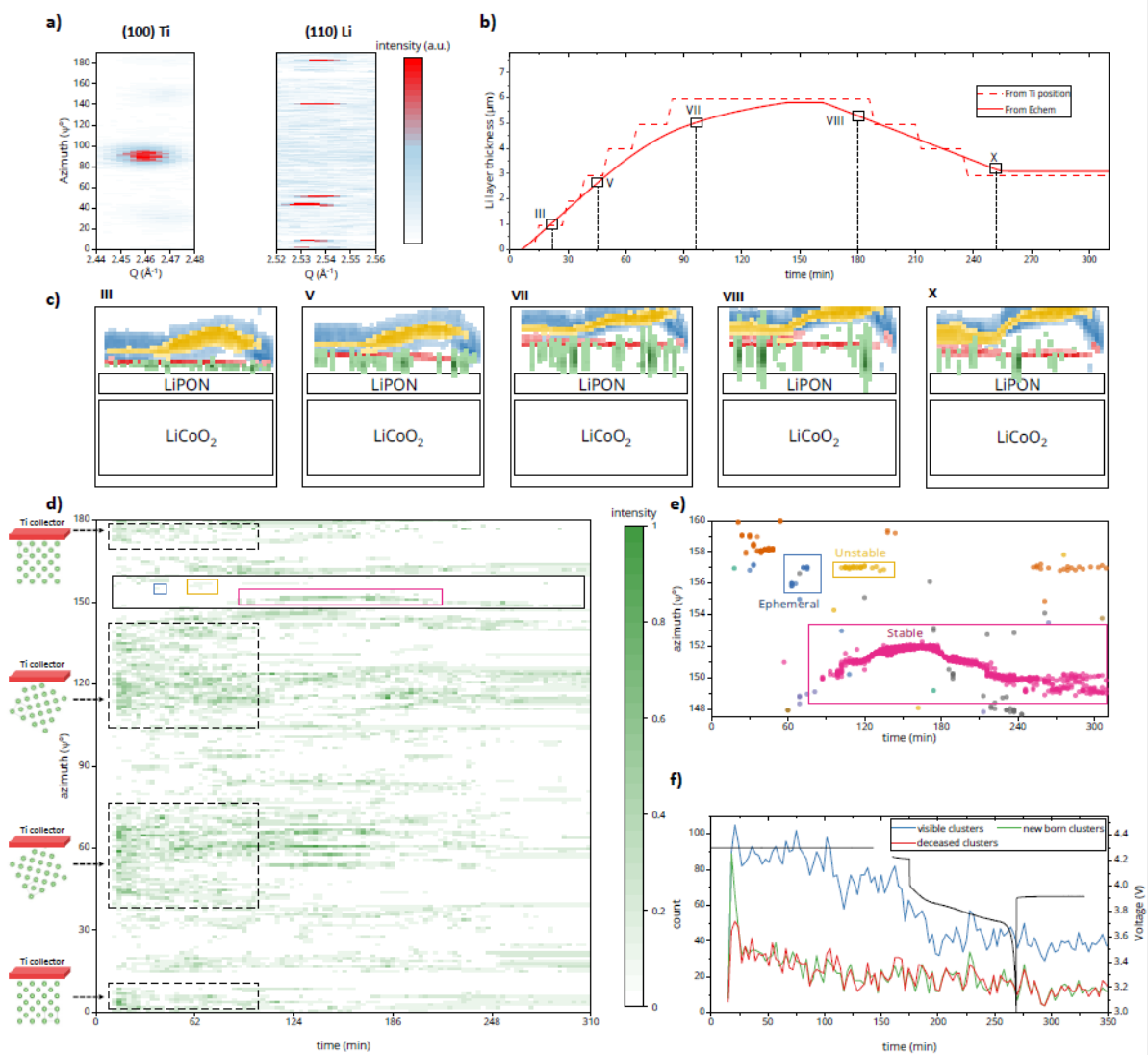

Figure 4: a) Typical detector images around the $(100)_{Ti}$ and $(110)_{Li}$ rings in the left and right panel, respectively. b) Li metal layer thickness calculated from the average Ti current position (dotted red) and from the electrochemistry (plain red line). c) Spatial maps of the Li, Ti, Au, and parylene positions during the charge and discharge shown in green, red, yellow and blue respectively. For each material, the colour corresponds to the normalized intensity of the $(110)_{Li}$, $(100)_{Ti}$, $(111)_{Au}$ and amorphous background of the parylene located around 1 Å$^{-1}$ [37]. Times, labelled as Latin numbers (III, VII, X etc.), are shown in b). Each pixel is 33 x 1 μm in the horizontal and vertical direction, respectively. Map size is 1.36 mm x 40 μm (horizontal x vertical). d) 2D plot of the azimuths of Li metal grains as a function of time. Color scale represents the normalized intensity of the diffraction spots. Black dotted rectangles are guide to the eyes showing the regions around 0°, 60°, 120° and 180° - azimuth of most of the Li metal grains. Black, blue, yellow and pink plain rectangles highlight the presence of 'ephemeral', 'unstable' and 'stable' Li metal grains better observed on the zoom in panel e). e) is a zoom of d). Each individual Li grain is given a different colour. As observed, some of the grains are only observed over a very short time lapse – so called 'Ephemeral' – while other grains are observable over longer times so called – 'unstable' and 'stable'. f) Total, new, 'deceased' number of grains in blue, green and red, respectively observed over the entire battery during the electrochemical sequence. black plain line is the voltage profile.

**Conclusion**

We have established fast operando μXRD imaging as a powerful tool to image Li intercalation and Li deposition as a function of grain orientation at the microscale, in a real cycling battery. This method was applied on an anode-free all solid state battery, the target device for high energy density. We found that fast delithiation of dense $LiCoO_2$ cathode is not limited by diffusion, while the full lithiation is limited by the formation of an insulating phase at the electrode surface. Design guidelines can be formulated from these direct observations, such as increasing the active area between electrolyte/electrode to tackle discharge capacity issues. At the anode side, we observed the deformation of the current collector, the Li/Ti epitaxial orientation relationship, the presence of large crystallographic coherent domains in Li metal (> 5 microns), and the fast dynamics of Li metal grains which occurs during plating/striping and rest. The results presented here highlight the potential of operando μXRD imaging to spatially resolve morphological/texture/cell parameter changes of most active and inactive cell components, at the microscale but with a 'field-of-view' up to the mm, hence giving statistical information on realistic battery. Fast operando μXRD imaging in 3D is also attainable now at 4[th] generation synchrotrons using advanced scattering tomographic modes. Critically, this method is not only limited to battery, but is applicable to a wide range of dynamic processes in crystals used in other energy related applications (fuel cells or solar cells), sensors, iontronics or electronic devices.

**Methods:**

**Fabrication of thin film batteries.** Eight-inch silicon wafers were used as substrates and maintained under a controlled atmosphere within an Ar-filled glovebox during sample transfer and storage operations. Thin-film batteries (TFBs) were fabricated in LETI cleanroom facilities using sequential layer deposition and patterning via UV photolithography and etching techniques. Film deposition steps were performed in an ENDURA PVD tool (Applied Materials) equipped with Ti, $LiCoO_2$, and $Li_3PO_4$ sputtering targets. SPR 220 photoresist was applied for all photolithography steps, with UV exposure performed using an MA8 mask aligner. Wet etching of $LiCoO_2$ and LiPON layers was conducted in a 1:1 $H_2SO_4:H_2O$ solution, while Pt layers were etched in a 1:3 $HCl:HNO_3$ solution. Ti top layers were etched in a $CHF_3$/Ar plasma using a Corial IL200 system. A polymer passivation layer was patterned, and a Ti redistribution layer was deposited as the negative current collector. Substrate thinning was performed through KOH etching. Finally, device singulation was achieved using laser ablation (Coherent GEM 100-$CO_2$ laser, λ = 10.6 μm, 20 W average power, 2 mm/s).

**SEM observation.** Cross section of thin film batteries was imaged with a MERLIN scanning electron microscope (ZEISS). The image was taken with a 45° tilted sample using 6 kV electron beam energy and 1.1 nA beam current. A 9 nm carbon layer was locally deposited to avoid electrostatic charging and improve the quality of the images.

**TFB mounting on PCB for operando characterization.** A specific printed circuit board (PCB) was designed and fabricated specifically for synchrotron operando characterization. Thin film batteries dies were attached using a UV cured epoxy. Wire bonding was carried out in order to electrically connect batteries to PCB pads.

**Electrochemical characterization of thin film batteries**. Electrochemical characterization of thin film batteries was carried out in Ar filled glove box before and after PCB mounting. Unless stated otherwise, electrochemical impedance spectroscopy (EIS), cyclic voltammetry (CV), and galvanostatic charge and discharge (GCPL) experiments were conducted at 25°C using a VMP3 multi-potentiostat (BioLogic) coupled with the EC-Lab software.

**Sample environment of electrochemical cycling during µXRD imaging experiment.** Ten microbatteries were positioned on a PCB. Before the experiment, a small current was applied to each battery to check electrical connection (5 nAh.cm$^{-2}$). Some of these batteries were imaged at open circuit voltage using µXRD imaging. The battery having the best image (alignment) and electrochemical test response was selected for operando characterisation. Battery were charged under potentiostatic conditions up to 4.3 V, while the current was measured. Discharge was performed at 4C with a current of 2 µA (corresponding to a current density of 0.4 mAh.cm²). Note that current shape obtained during the synchrotron experiment is very similar to previous test in the lab as shown Fig. S12.

**µXRD imaging experiment.** µXRD mapping was performed on ID13 at the European Synchrotron Radiation Facility (ESRF) using a 1 x 1 µm beam (vertical and horizontal directions, z and x) at 18 keV. The battery was 40 microns thick x 2 mm wide x 350 µm long (z, x, y directions, respectively). The cathode, electrolyte, current collector layer stacking was oriented along the z direction. The battery was positioned so that the X-ray beam crosses the battery along the y-axis. During the acquisition, the battery was moved horizontally (x) and vertically (z) to scan the entire electrode stack, producing (x, z) maps of 49 x 49 pixels with pixel size of 33 x 1 µm in approx. 5 min. The counting time at each position is 0.05 sec (2 min of counting time per map and approx. 3 min of motor movement). "Horizontal fly scans" were performed in which the battery was continuously moved in the x direction from 0 to 1.6 mm while the shutter remained opened. Detector images were averaged over 33 µm of battery displacement. Therefore, each detector image is an average in the x direction of 33 µm of the sample. After every horizontal continuous scan, the battery was displaced vertically by 1 µm (in z direction) and a new continuous scan in x was performed until the full 2D maps were produced. This continuous scanning reduces a lot the dose, it spreads over the 33 µm scanned horizontally. WAXS patterns were recorded using a Dectris EigerX 4M (2070 x 2167 pixel, 75 micron pixel size). Measured Q range is 0.96 – 6.09 Å$^{-1}$. After the sample, the direct beam hits a beam stop approximately at the center of the detector, therefore, full Debye Scherer rings can potentially be measured with however some shadowing effects coming from the sample as shown Fig. S11. Detector calibration was performed with a reference sample $Al_2O_3$ using PyFAI[38]. Sample-to-detector distance calibration was corrected using the Ti current collector peaks.

**µXRD imaging data analysis.** Detector images are azimuthally integrated using PYFAI transforming the detector image (in pixels) in a map showing the intensity as a function of Q and azimuth (x and y axis). From these maps, the rest of the analysis were performed using a homemade python based code. LCO: First, the intensity ratio between LCO$^\perp$ and LCO$^\parallel$ was determined on the pristine battery (Fig. 1g). For this intensity stripes for (003)$_{LCO}$ peaks at 90° (LCO$^\parallel$) and 160° (LCO$^\perp$) were integrated over a +/- 10° azimuth range to obtained a 'classical' diffraction peak (intensity as a function of Q). Gaussian peak fitting is performed to estimate the peak areas. To obtain the cell parameter evolution as a function of time and orientation (Fig 2. and Fig. 3), a similar approach was performed on the (003)$_{LCO}$ and (110)$_{LCO}$ for all LCO pixels in each maps (approx. 96000 detector image treated). Peak position was obtained from the Gaussian peak position. To estimate Li concentration from peak position, a calibration curve was extracted from the literature[24]. Li metal analysis: The analysis was conducted on the (110)$_{Li}$ peak, and hence all azimuthally integrated detector images were reduced focusing on the data in the 2.53 – 2.55 Å$^{-1}$ range. The typical data manipulated is a 5 dimensional array (100 time steps, 49 y direction step, 49 z direction steps, 360 azimuths, 10 pts in between 2.53 – 2.55 Å$^{-1}$) of the detector intensity. The next step consist in removing the background due total scattering. This is performed by removing a detector image free of Li metal peaks for each time, y direction and z direction. Then, Li metal peaks on the detector are segmented by tresholding the intensity with 10 times the mean intensity of the detector image in the reduced Q range around the (110)$_{Li}$ which is mostly composed of the remaining contribution from the background. Finally, from the segmented dataset, the number of Li clusters, their size, azimuth and duration are determined. All this data analysis process is performed using a homemade Python code using mostly numpy packages.

**Li concentration modelling.** Lithium concentration is modelled in COMSOL with a variable diffusion coefficient as described in Celè et al[7]. 1D Drift-diffusion equation is used for both LiCoO$_2$ and LiPON domain, while Butler-Volmer equation is used to model the cathode/electrolyte interface. All equations are casted in weak form and directly implemented in the COMSOL weak-PDE model interface to be numerically solved.

**Acknowledgement :** Authors thank F. Monaco for his help during beamtime. Beamtime at the ESRF was granted within the Battery Pilot Hub MA4929 "Multi-scale Multi-techniques investigations of Li-ion batteries: towards a European Battery Hub".

**Author contributions:** The study was initiated as a collaboration between the Institut of Interdisciplinary Research (IRIG) and the LETI at a meeting between S.L. and S.O. J.C. and S.O. proposed the investigated system and exposed characterization needs towards modelling validation. S.L. proposed the x-ray scanning microbeam method to obtain concentration gradients across the thin film battery, selected the beamline and obtained the beamtime. S.O. selected the test vehicle and obtained the cleanroom fabrication. J.C., Q.J., S.T., S.O., and S.L. conceptualized the study. The LETI team (S.O., J.C., L.C.) designed, fabricated and tested the devices and dedicated operando printed circuit board (PCB). The IRIG team (Q.J., S.T., S.L.), with assistance from J.C. and L.C., conducted the synchrotron experiments, applying methods initially developed by S.T. and S.L. on ID13. The ID13 beamline was set up by M.B. and A.M., who assisted with sample alignment and data acquisition. Q.J. performed a preliminary analysis of the integrated XRD data. J.C. refined the analysis to obtain LCO phase transitions and spatial heterogeneities, with input from Q.J., S.T., and S.P.; Q.J. designed and performed texture imaging analysis on lithium metal anode. All authors interpreted and discussed the data. Q.J. and J.C. wrote the initial draft, which was revised by S.L., S.O., and S.T. All authors contributed to further manuscript revisions.

**Data Availability :** Data collected at the European Synchrotron Radiation Facility are accessible under the DOI : doi.org/10.15151/ESRF-DC-2003355775

**Code availability**: All code used in this work is available from the corresponding authors upon reasonable request.

**References:**

(1) Janek, J.; Zeier, W. G. Challenges in Speeding up Solid-State Battery Development. *Nat Energy* **2023**, *8* (3), 230–240. https://doi.org/10.1038/s41560-023-01208-9.
(2) Li, M.; Liu, T.; Shi, Z.; Xue, W.; Hu, Y.; Li, H.; Huang, X.; Li, J.; Suo, L.; Chen, L. Dense All-Electrochem-Active Electrodes for All-Solid-State Lithium Batteries. *Advanced Materials* **2021**, *33* (26), 2008723. https://doi.org/10.1002/adma.202008723.

(3) Wang, M. J.; Carmona, E.; Gupta, A.; Albertus, P.; Sakamoto, J. Enabling "Lithium-Free" Manufacturing of Pure Lithium Metal Solid-State Batteries through in Situ Plating. *Nat Commun* **2020**, *11* (1), 5201. https://doi.org/10.1038/s41467-020-19004-4.

(4) Nanda, S.; Gupta, A.; Manthiram, A. Anode-Free Full Cells: A Pathway to High-Energy Density Lithium-Metal Batteries. *Advanced Energy Materials* **2021**, *11* (2), 2000804. https://doi.org/10.1002/aenm.202000804.

(5) Kim, J. Y.; Park, J.; Lee, M. J.; Kang, S. H.; Shin, D. O.; Oh, J.; Kim, J.; Kim, K. M.; Lee, Y.-G.; Lee, Y. M. Diffusion-Dependent Graphite Electrode for All-Solid-State Batteries with Extremely High Energy Density. *ACS Energy Lett.* **2020**, *5* (9), 2995–3004. https://doi.org/10.1021/acsenergylett.0c01628.

(6) Johnson, A. C.; Dunlop, A. J.; Kohlmeyer, R. R.; Kiggins, C. T.; Blake, A. J.; Singh, S. V.; Beale, E. M.; Zahiri, B.; Patra, A.; Yue, X.; Cook, J. B.; Braun, P. V.; Pikul, J. H. Strategies for Approaching One Hundred Percent Dense Lithium-Ion Battery Cathodes. *Journal of Power Sources* **2022**, *532*, 231359. https://doi.org/10.1016/j.jpowsour.2022.231359.

(7) Celè, J.; Franger, S.; Lamy, Y.; Oukassi, S. Minimal Architecture Lithium Batteries: Toward High Energy Density Storage Solutions. *Small* **2023**, *19* (16), 2207657. https://doi.org/10.1002/smll.202207657.

(8) Li, Y.; El Gabaly, F.; Ferguson, T. R.; Smith, R. B.; Bartelt, N. C.; Sugar, J. D.; Fenton, K. R.; Cogswell, D. A.; Kilcoyne, A. L. D.; Tyliszczak, T.; Bazant, M. Z.; Chueh, W. C. Current-Induced Transition from Particle-by-Particle to Concurrent Intercalation in Phase-Separating Battery Electrodes. *Nature Mater* **2014**, *13* (12), 1149–1156. https://doi.org/10.1038/nmat4084.

(9) Kasemchainan, J.; Zekoll, S.; Spencer Jolly, D.; Ning, Z.; Hartley, G. O.; Marrow, J.; Bruce, P. G. Critical Stripping Current Leads to Dendrite Formation on Plating in Lithium Anode Solid Electrolyte Cells. *Nat. Mater.* **2019**, *18* (10), 1105–1111. https://doi.org/10.1038/s41563-019-0438-9.

(10) Zheng, J.; Kim, M. S.; Tu, Z.; Choudhury, S.; Tang, T.; Archer, L. A. Regulating Electrodeposition Morphology of Lithium: Towards Commercially Relevant Secondary Li Metal Batteries. *Chem. Soc. Rev.* **2020**, *49* (9), 2701–2750. https://doi.org/10.1039/C9CS00883G.

(11) Sanchez, A. J.; Dasgupta, N. P. Lithium Metal Anodes: Advancing Our Mechanistic Understanding of Cycling Phenomena in Liquid and Solid Electrolytes. *J. Am. Chem. Soc.* **2024**, *146* (7), 4282–4300. https://doi.org/10.1021/jacs.3c05715.

(12) Strauss, F.; Kitsche, D.; Ma, Y.; Teo, J. H.; Goonetilleke, D.; Janek, J.; Bianchini, M.; Brezesinski, T. Operando Characterization Techniques for All-Solid-State Lithium-Ion Batteries. *Advanced Energy and Sustainability Research* **2021**, *2* (6), 2100004. https://doi.org/10.1002/aesr.202100004.

(13) Kazyak, E.; Wang, M. J.; Lee, K.; Yadavalli, S.; Sanchez, A. J.; Thouless, M. D.; Sakamoto, J.; Dasgupta, N. P. Understanding the Electro-Chemo-Mechanics of Li Plating in Anode-Free Solid-State Batteries with Operando 3D Microscopy. *Matter* **2022**, *5* (11), 3912–3934. https://doi.org/10.1016/j.matt.2022.07.020.

(14) Otoyama, M.; Sakuda, A.; Hayashi, A.; Tatsumisago, M. Optical Microscopic Observation of Graphite Composite Negative Electrodes in All-Solid-State Lithium Batteries. *Solid State Ionics* **2018**, *323*, 123–129. https://doi.org/10.1016/j.ssi.2018.04.023.

(15) Perrenot, P.; Bayle-Guillemaud, P.; Jouneau, P.-H.; Boulineau, A.; Villevieille, C. Operando Focused Ion Beam–Scanning Electron Microscope (FIB-SEM) Revealing Microstructural and Morphological Evolution in a Solid-State Battery. *ACS Energy Lett.* **2024**, *9* (8), 3835–3840. https://doi.org/10.1021/acsenergylett.4c01750.

(16) Nomura, Y.; Yamamoto, K.; Fujii, M.; Hirayama, T.; Igaki, E.; Saitoh, K. Dynamic Imaging of Lithium in Solid-State Batteries by Operando Electron Energy-Loss Spectroscopy with Sparse Coding. *Nat Commun* **2020**, *11* (1), 2824. https://doi.org/10.1038/s41467-020-16622-w.

(17) Motoyama, M.; Ejiri, M.; Yamamoto, T.; Iriyama, Y. In Situ Scanning Electron Microscope Observations of Li Plating/Stripping Reactions with Pt Current Collectors on LiPON Electrolyte. *J. Electrochem. Soc.* **2018**, *165* (7), A1338–A1347. https://doi.org/10.1149/2.0411807jes.

(18) Wu, X.; Billaud, J.; Jerjen, I.; Marone, F.; Ishihara, Y.; Adachi, M.; Adachi, Y.; Villevieille, C.; Kato, Y. Operando Visualization of Morphological Dynamics in All-Solid-State Batteries. *Advanced Energy Materials* **2019**, *9* (34), 1901547. https://doi.org/10.1002/aenm.201901547.

(19) Lewis, J. A.; Cortes, F. J. Q.; Liu, Y.; Miers, J. C.; Verma, A.; Vishnugopi, B. S.; Tippens, J.; Prakash, D.; Marchese, T. S.; Han, S. Y.; Lee, C.; Shetty, P. P.; Lee, H.-W.; Shevchenko, P.; De Carlo, F.; Saldana, C.; Mukherjee, P. P.; McDowell, M. T. Linking Void and Interphase Evolution to Electrochemistry in Solid-State Batteries Using Operando X-Ray Tomography. *Nat. Mater.* **2021**, *20* (4), 503–510. https://doi.org/10.1038/s41563-020-00903-2.

(20) Colalongo, M.; Vostrov, N.; Martens, I.; Zatterin, E.; Richard, M.-I.; Cadiou, F.; Jacquet, Q.; Drnec, J.; Leake, S. J.; Kallio, T.; Zhu, X.; Lyonnard, S.; Schulli, T. Imaging Inter - and Intra-Particle Features in Crystalline Cathode Materials for Li-Ion Batteries Using Nano-Focused Beam Techniques at 4th Generation Synchrotron Sources. *Microstructures* **2024**, *4* (4). https://doi.org/10.20517/microstructures.2024.19.

(21) Berhaut, C. L.; Mirolo, M.; Dominguez, D. Z.; Martens, I.; Pouget, S.; Herlin-Boime, N.; Chandesris, M.; Tardif, S.; Drnec, J.; Lyonnard, S. Charge Dynamics Induced by Lithiation Heterogeneity in Silicon-Graphite Composite Anodes. *Advanced Energy Materials* **2023**, *13* (44), 2301874. https://doi.org/10.1002/aenm.202301874.

(22) Martens, I.; Vostrov, N.; Mirolo, M.; Leake, S. J.; Zatterin, E.; Zhu, X.; Wang, L.; Drnec, J.; Richard, M.-I.; Schulli, T. U. Defects and Nanostrain Gradients Control Phase Transition Mechanisms in Single Crystal High-Voltage Lithium Spinel. *Nat Commun* **2023**, *14* (1), 6975. https://doi.org/10.1038/s41467-023-42285-4.

(23) Jacquet, Q.; Profatilova, I.; Baggetto, L.; Alrifai, B.; Addes, E.; Chassagne, P.; Blanc, N.; Tardif, S.; Daniel, L.; Lyonnard, S. Mapping Reaction Mechanism During Overcharge of a LiNiO2/Graphite–Silicon Lithium-Ion Battery: A Correlative Operando Approach by Simultaneous Gas Analysis and Synchrotron Scattering Techniques. *Advanced Energy Materials* n/a (n/a), 2404080. https://doi.org/10.1002/aenm.202404080.

(24) Reimers, J. N.; Dahn, J. R. Electrochemical and In Situ X-Ray Diffraction Studies of Lithium Intercalation in Li x CoO2. *J. Electrochem. Soc.* **1992**, *139* (8), 2091. https://doi.org/10.1149/1.2221184.

(25) Mukai, K.; Uyama, T.; Nonaka, T. Revisiting LiCoO2 Using a State-of-the-Art In Operando Technique. *Inorg. Chem.* **2020**, *59* (15), 11113–11121. https://doi.org/10.1021/acs.inorgchem.0c01598.

(26) Merryweather, A. J.; Schnedermann, C.; Jacquet, Q.; Grey, C. P.; Rao, A. Operando Optical Tracking of Single-Particle Ion Dynamics in Batteries. *Nature* **2021**, *594* (7864), 522–528. https://doi.org/10.1038/s41586-021-03584-2.

(27) Dudney, N. J.; Jang, Y.-I. Analysis of Thin-Film Lithium Batteries with Cathodes of 50 Nm to 4 Mm Thick LiCoO2. *Journal of Power Sources* **2003**, *119–121*, 300–304. https://doi.org/10.1016/S0378-7753(03)00162-9.

(28) Matsuda, Y.; Kuwata, N.; Kawamura, J. Thin-Film Lithium Batteries with 0.3–30 Mm Thick LiCoO2 Films Fabricated by High-Rate Pulsed Laser Deposition. *Solid State Ionics* **2018**, *320*, 38–44. https://doi.org/10.1016/j.ssi.2018.02.024.

(29) Grenier, A.; Reeves, P. J.; Liu, H.; Seymour, I. D.; Märker, K.; Wiaderek, K. M.; Chupas, P. J.; Grey, C. P.; Chapman, K. W. Intrinsic Kinetic Limitations in Substituted Lithium-Layered Transition-Metal Oxide Electrodes. *J. Am. Chem. Soc.* **2020**, *142* (15), 7001–7011. https://doi.org/10.1021/jacs.9b13551.

(30) Choi, J.; Son, B.; Ryou, M.-H.; Kim, S. H.; Ko, J. M.; Lee, Y. M. Effect of LiCoO2 Cathode Density and Thickness on Electrochemical Performance of Lithium-Ion Batteries. *J. Electrochem. Sci. Technol* **2013**, *4* (1), 27–33. https://doi.org/10.33961/JECST.2013.4.1.27.

(31) Celè, J. Thin Film Batteries for Integrated High Energy Density Storage. These de doctorat, université Paris-Saclay, 2023. https://theses.fr/2023UPASF095 (accessed 2024-12-10).

(32) Zhang, M.-X.; Kelly, P. M. Edge-to-Edge Matching and Its Applications: Part I. Application to the Simple HCP/BCC System. *Acta Materialia* **2005**, *53* (4), 1073–1084. https://doi.org/10.1016/j.actamat.2004.11.007.

(33) Burgers, W. G. On the Process of Transition of the Cubic-Body-Centered Modification into the Hexagonal-Close-Packed Modification of Zirconium. *Physica* **1934**, *1* (7–12), 561–586. https://doi.org/10.1016/S0031-8914(34)80244-3.

(34) Wang, W.; Li, Z.; Chen, S.; Wang, Y.; He, Y.-S.; Ma, Z.-F.; Li, L. The Overlooked Role of Copper Surface Texture in Electrodeposition of Lithium Revealed by Electron Backscatter Diffraction. *ACS Energy Lett.* **2024**, *9* (1), 168–175. https://doi.org/10.1021/acsenergylett.3c02132.

(35) Motoyama, M.; Ejiri, M.; Iriyama, Y. Modeling the Nucleation and Growth of Li at Metal Current Collector/LiPON Interfaces. *J. Electrochem. Soc.* **2015**, *162* (13), A7067. https://doi.org/10.1149/2.0051513jes.

(36) Li, X.; Chen, C.; Fu, Z.; Wang, J.; Hu, C. Compression Promotes the Formation of {110} Textures during Homoepitaxial Deposition of Lithium. *Energy Storage Materials* **2023**, *58*, 155–164. https://doi.org/10.1016/j.ensm.2023.03.019.

(37) Lee, J.-H.; Kim, A. Structural and Thermal Characteristics of the Fast-Deposited Parylene Substrate for Ultra-Thin Organic Light Emitting Diodes. *Organic Electronics* **2017**, *47*, 147–151. https://doi.org/10.1016/j.orgel.2017.05.005.

(38) Kieffer, J.; Karkoulis, D. PyFAI, a Versatile Library for Azimuthal Regrouping. *J. Phys.: Conf. Ser.* **2013**, *425* (20), 202012. https://doi.org/10.1088/1742-6596/425/20/202012.

**Supporting information**

**Operando imaging of crystal structure and orientation in all components of all-solid-state-batteries**

Quentin Jacquet[1#], Jacopo Cele[2, 3#], Lara Casiez[2], Samuel Tardif[4], Asma Medjaheh[5], Stephanie Pouget[4], Manfred Burghammer[5], Sandrine Lyonnard[1]*, Sami Oukassi[2]*

[1] Univ. Grenoble Alpes, CEA, CNRS, Grenoble INP, IRIG, SyMMES, 38000 Grenoble, France.

[2] Univ. Grenoble Alpes, CEA, Leti Grenoble F-38000, France

[3] ICMMO (UMR CNRS 8182) Univ. Paris-Sud Univ. Paris-Saclay Orsay 91190, France,

[4] University Grenoble Alpes, CEA, CNRS, IRIG, MEM, 38000 Grenoble, France.

[5] European Synchrotron Radiation Facility (ESRF), CS 40220, 71 Avenue des Martyrs, 38043 Grenoble, France

Corresponding authors: sandrine.lyonnard@cea.fr; sami.oukassi@cea.fr; quentin.jacquet@cea.fr; Jacopo.cele@cea.fr

# Equal contribution

## 1-Electrochemical behaviour

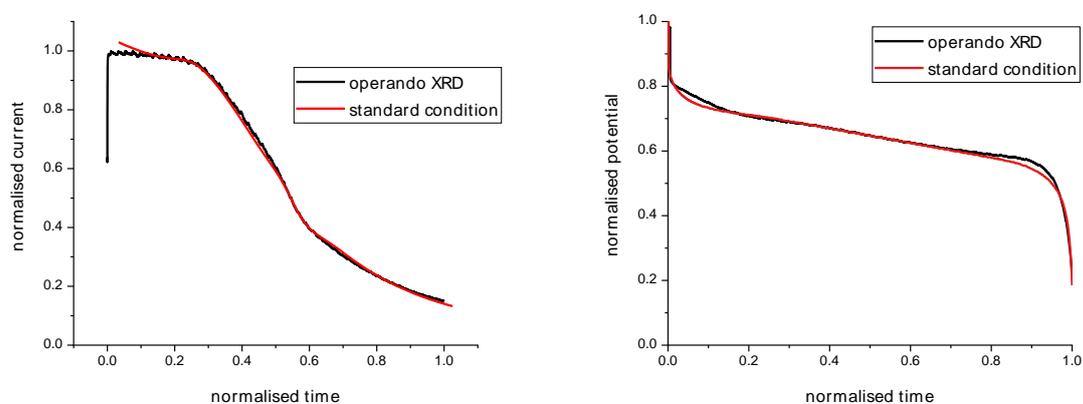

**Figure S1. Comparison of the operando cell performance with other another cell cycled in the lab:** The charge and discharge current and voltage profile are normalised in order to simplify the comparison. The cell cycled during the operando XRD experiment present the same profiles as a cell tested in lab condition under controlled atmosphere. No evident interference of the environment of the test setup are identified.

## 2-Peak Fitting

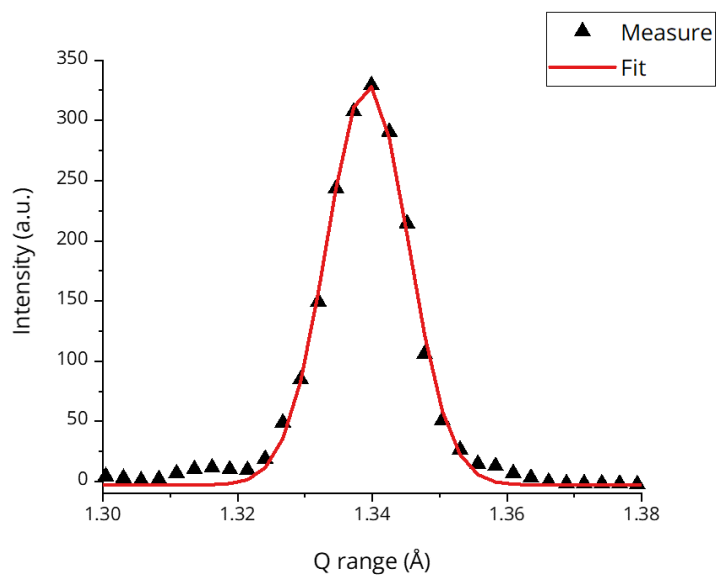

**Figure S2. Typical fit of the (003) peaks used to determine the c lattice parameter evolution.** Raw data integrate along the azimuthal axis, and Gaussian fitting to refine the (003) peak position for one single pixel.

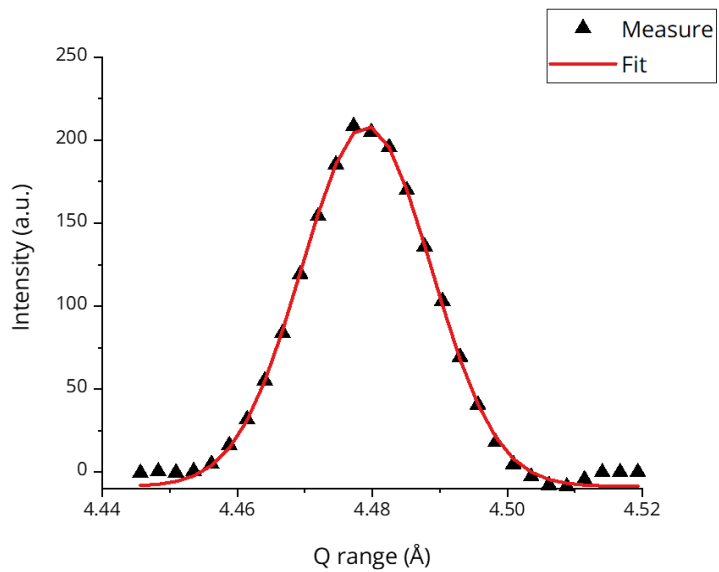

**Figure S3. Typical fit of (110) peaks used to determined the a lattice parameter variation.** Raw data integrate along the azimuthal axis, and Gaussian fitting to refine the (110) peak position for one single pixel.

**3-Discussion on the origin of different c lattice evolution between LCO$^{\perp}$ and LCO$^{\parallel}$.**

First, we rule out any detector issues.

1 – Detector calibration. During the experiment, the detector is not perfectly perpendicular to the incoming beam direction. As a consequence, measured Debye Scherrer rings are actually oval. To correct this, detector needs to be calibrated. Calibration corresponds to the determination of the sample to detector distance together with all the angles of the detector. To perform detector calibration, a reference sample ($Al_2O_3$ in this case) is measured in the same conditions as the sample. Well-known diffraction rings of $Al_2O_3$ are fitted changing the position/angle of the detector. After performing the calibration, raw detector images are corrected by obtain proper Debye Scherrer rings. To verify the quality of the calibration, diffraction intensity maps with the azimuth (y direction) and Q range (x direction) are plotted. Looking at the (111)$_{Au}$ as a function of azimuth shown in the Figure Sa below, one can see that the peak position is the same for all azimuths showing that the calibration is correct.

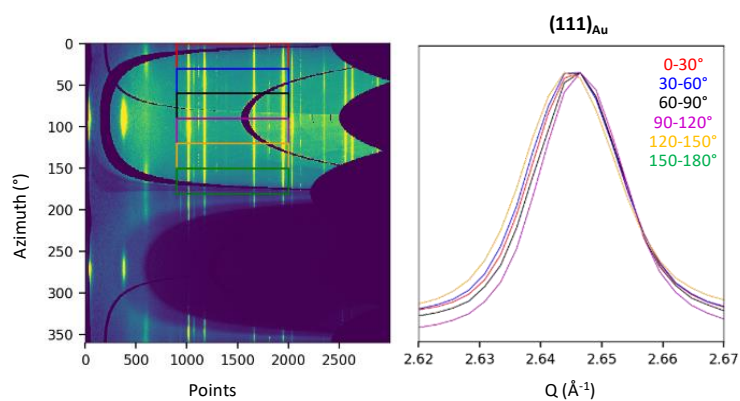

**Figure Sa:** left panel is the result of an azimuthal integration of a detector image collected at the Au RDL position in the battery. On the right panel is the integrated diffraction pattern for diffraction azimuths zoomed on the $(111)_{Au}$. There is almost no shift between the diffraction signal coming from the different azimuth showing that the calibration is correct.

2 – <u>Effect of Detector hole.</u> Peak position can be influenced by holes in the detector. Indeed, if detector holes are present at the peak position, the shape of the intensity profile can be distorted, and hence the position of the peak might change. Indeed, some of the (003) spot intensity are absent due to detector hole. However, it is not the case of the intensity that we have used for the data analysis, hence ruling out the effect of detector hole on the (003) peak position.

Having ruled out possible unphysical origin for the discrepancy by c lattice parameters of **LCO$^\perp$ and LCO$^\parallel$**. We now turn to physical or chemical origin.

4 - <u>Mechanical effects.</u> In the following, we elaborate on possible mechanical effect. Lattice parameter variation during lithiation is approx. 0.1% and 1% for a and c, respectively. Therefore, in the following we will focus on the mechanical effects due to c lattice expansion. Interlayer expansion (c increase) in LCO$^\perp$ leads to crystal growth (xy) plane while LCO$^\parallel$ expansion is out of plane (z direction). LCO crystal have a columnar morphology. They are long in the out of plane direction (approx. 10 µm), and short in the in plane direction (approx. 100 nm). 1% c lattice expansion leads to 100 nm crystal expansion out of plane for LCO$^\parallel$ crystals and 1 nm expansion in plane for LCO$^\perp$. From the cross section SEM shown Fig. 1b some porosity can be observed between the grains and it is likely that 1 nm expansion could be possible without drastic mechanical compression. Expansion of 100 nm in the out of plane direction could be more difficult hence leading to strong stress. One possible explanation for the modified phase diagram for LCO$^\parallel$ could be compression during expansion.

5 - <u>Chemical effects.</u> LCO could have defects. Typical defects include extra Li and/or Co in the lithium layer. However, the presence of defects in known to suppress the biphasic phase transition, which is replaced by a solid solution reaction (constant evolution of the lattice parameter during delithiation). For both **LCO$^\perp$ and LCO$^\parallel$**, we observe a biphasic transition hence ruling out potential chemical heterogeneity with orientation.

## 4 – Determination of Li concentration

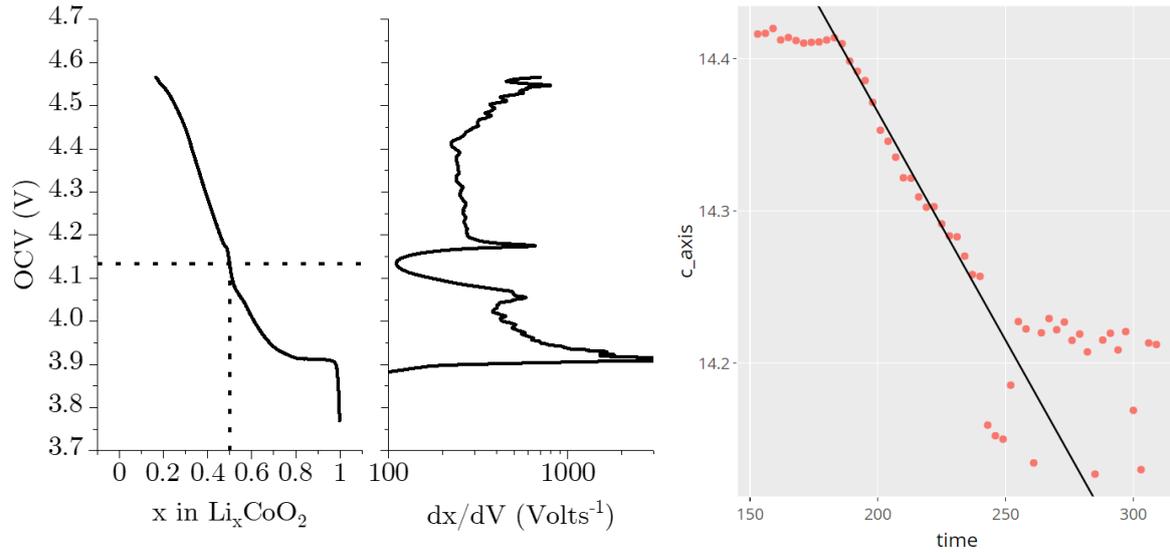

**Figure S4. Calibration of Li concentration vs electrochemical curve. a)** In order to determine a relationship between the OCV and the fraction of intercalated lithium (x in $Li_xCoO2$), the order-disorder transition at x=0.5 must be identified. By plotting the inverse derivative of the voltage with respect to the charge fraction (right), the order-disorder transition voltage is identified by the minimum of the curve. The total charge to reach the transition is fixed equal to half of the total lithium available. The resulting calibration is shown on the left. **b)** linear variation of the c axis parameter during time. Since the current is constant and the intercalation homogeneous across the all cathode, a direct relationship between lithium intercalation fraction and c axis parameters can be defined.

**Calibration of Lithium concentration x versus c axis parameter**

According to literature, a linear expansion of the c axis is expected with respect to the quantity of intercalated lithium for the metallic phase. The behaviour is also confirmed during the constant current discharge of the present work (Figure S4 b). The calibration of the c axis parameter with respect to the concentration of lithium is an empirical linear equation. The two points used to determine the slope and intercept of the equation is the final value of the charge and the final point of the discharge. For the charge voltage $4.2\ V$ we can determine the intercalation fraction $x_{li} = 0.45$ and the corresponding c axis parameter $c_{axis} = 14.43$ Å using the calibrated OCV. At the end of the discharge we have the final intercalation fraction $x_{li} = 0.75$ and the corresponding c axis parameter $c_{axis} = 14.21$ Å. We obtain:

$$x_{li}(t) = -(c_{axis}(t) + 1 - c_{axis}(t=0)) * \frac{0.3}{0.22} = 20.13 - c_{axis}(t) * \frac{0.3}{0.22}$$

This first equation is valid only for the metallic phase II (i.e. $x_{li} < 0.75$). For values above this threshold, the phase transition takes place, and two phases exists at the same time. Assuming no change in diffraction intensity and no change in the $c_{axis}$ during the phase transition, it is still possible to define a linear relationship between the mean value of the $c_{axis}$ parameters $\bar{c}_{axis}$ and the mean intercalation fraction $\bar{x}_{li}$: $\bar{x}_{li}$ is proportional to the mean $\bar{c}_{axis}$ which is the linear combination of the two phase's $c_{axis}$ and the weights are the phase fractions:

:

$$\bar{c}_{axis}(t) = \alpha(t) * c_{phase\_I} + \beta(t) * c_{phase\_II} \propto \bar{x}_{li}(t)$$

With $\alpha$ and $\beta$ the phase fractions:

$$\alpha(t) + \beta(t) = 1$$

In particular, assuming the only possible value of $x_{li}$ are the phases extrema ($x_{li} = 0.75$ and $x_{li} = 1$), we can write $\bar{x}_{li}$ as:

$$\bar{x}_{li}(t) = \alpha(t) * 1 + \beta(t) * 0.75$$

and

Recalling $x_{li} = 0.75$ for $c_{axis} = 14.21$ Å and $c_{axis} = 14.03$ Å for $x_{li} = 1$ we have the following system

$$\begin{cases} \bar{x}_{li}(t) = \alpha(t) * 1 + \beta(t) * 0.75 \\ \bar{c}_{axis}(t) = \alpha(t) * 14.03 + \beta(t) * 14.21 \\ \alpha(t) + \beta(t) = 1 \end{cases}$$

Solving for $\bar{x}_{li}(t)$ and $\bar{c}_{axis}(t)$:

$$\frac{\bar{x}_{li}(t) - 0.75}{\bar{c}_{axis}(t) - 14.21} = \frac{\alpha(t) * 0.25}{-\alpha(t) * 0.18}$$

We obtain

$$\begin{cases} \bar{x}_{li}(t) = 20.49 - \frac{0.25}{0.18} \bar{c}_{axis}(t) & x_{li} \geq 0.75 \\ \bar{x}_{li}(t) = 20.13 - \frac{0.3}{0.22} \bar{c}_{axis}(t) & x_{li} < 0.75 \end{cases}$$

## 5 – Ti texture

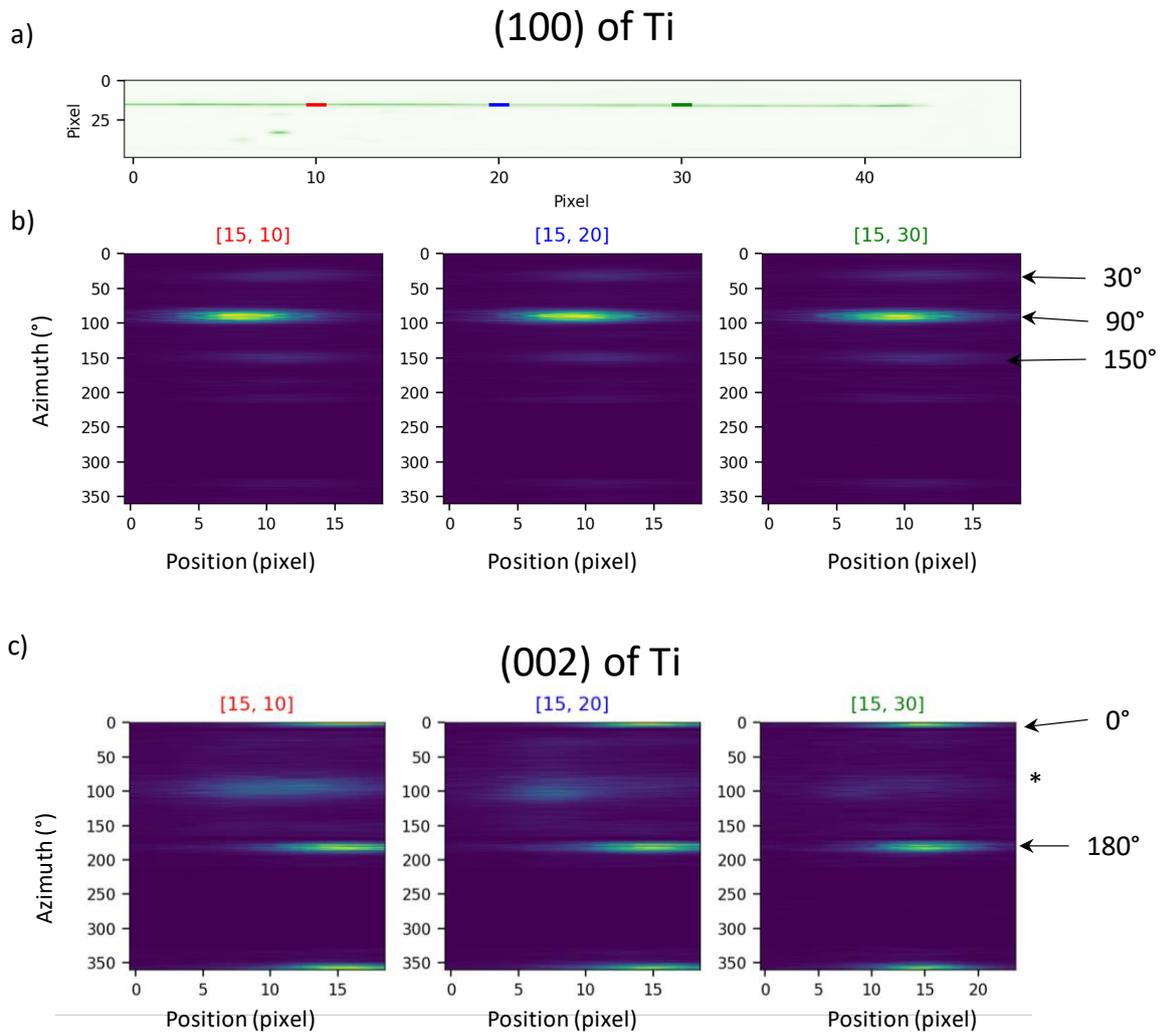

**Figure S5:** a) Spatial map of the Ti current collector $(100)_{Ti}$ signal. Pixel size is 1 x 33 microns (vertical x horizontal directions). Three pixels are selected at different horizontal positions, namely, 10, 15, and 20 in red, blue, and green respectively. b) Azimuthal integration of the detector image for the three positions shown in a). Vertical and horizontal axis corresponds to the azimuth and pixel on the detector, respectively. Pixel on the detector can be converted into Q range (varying in this case between 2.44 to 2.48 Å$^{-1}$). On all detector image, an intense spot at 90° is observed together with lower intensity spots at 30° and 150°. c) Azimuthal integration of the detector image for the three positions shown in a) zoomed on the $(002)_{Ti}$. Vertical and horizontal axis corresponds to the azimuth and pixel on the detector, respectively.

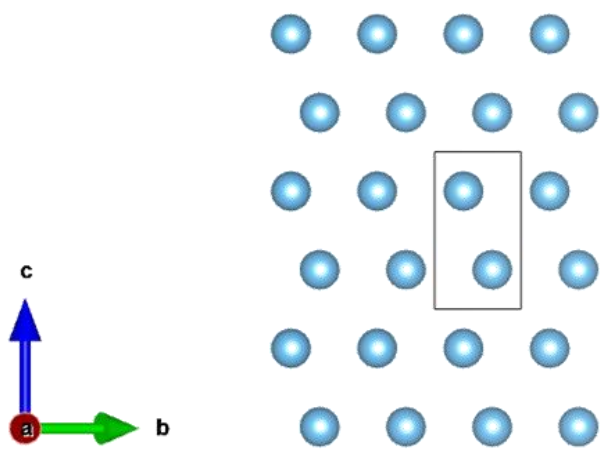

**Figure S6.** Top view of the Ti current collector (*a* lattice parameter pointing towards us since the (100)$_{Ti}$ since the peak is observed at azimuth 90°

## 6 – Li texture

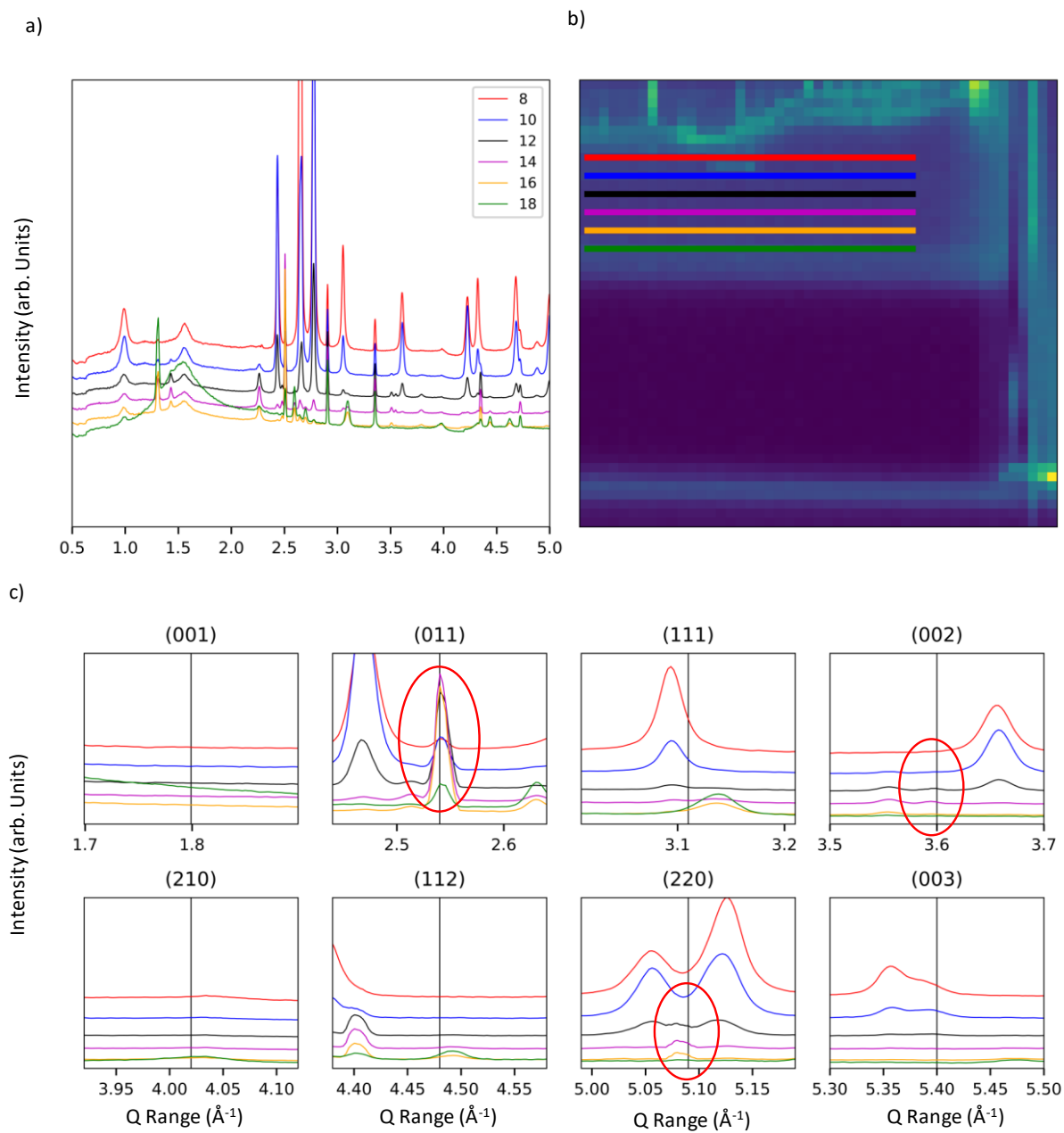

**Figure S7. Relative intensity of Li metal peaks.** a) Line-wise averaged diffraction patterns. Lines were selected in the Li metal electrode. Each line is represented in the X-ray transmission image in b). c) zoom of the diffraction pattern where Li metal peaks are located. (011)$_{Li}$ diffraction peaks is quite clearly visible. Most of the other peaks are invisible but (002)$_{Li}$ and (220)$_{Li}$.

## 7 - Discussion on the orientation relationship between Li and Ti.

<u>What would be the crystal orientation relationship between Li and Ti?</u> Li crystallize in Body Centered Cubic (BCC) structure with a = 3.491 Å and Ti features a Hexagonal Close Packed (HCP) structure with a = 2.951 Å and c = 4.684 Å. BCC and HCP orientation relationship (OR) have been measured experimentally for a wide range of materials and four different ORs have been observed: the Burger OR [1], The Potter OR[2], The Pistch–Schrader OR[3], The Rong–Dunlop (R–D) OR[4]. Several theories have been proposed to explain the ORs depending on the nature of the two phases in presence. Amongst them, the edge-edge theory proposed by Zhang-Kelly successfully predicts the OR depending on the cell parameter ratio of both phases[3]. To determine possible OR, matching directions and planes are assumed to only be closed-packed directions and planes. Directions and planes are matching when interatomic spacing misfits and the misfit between d lattice spacings are below 6% and 10% respectively. Starting with the interatomic spacings, they can be estimated based on the $a_H/a_B$ ratio which is equal to 0.845 for Ti/Li metals. For this value of $a_H/a_B$ Zhang et al. predicted that only the <11-20>$_H$/<111>$_B$ and <10-10>$_H$/<113>$_B$ match the < 6% criteria. Regarding the lattice plane misfit, the close packed planes for HCP and BCC are {10-11}$_H$, {10-10}$_H$, {0002}$_H$ and {110}$_B$, {200}$_B$, {111}$_B$. Calculating the mismatch between the different planes, {0002}$_H$/{110}$_B$ d spacing ratio is 5.1 % and it is the only pair with a mismatch inferior to 10%. Possible ORs between Ti and Li are hence <11-20>$_H$/<111>$_B$ with {0002}$_H$/{110}$_B$ which is the Burger OR ([2-1-10]$_{Ti}$//[1-11]$_{Li}$ with (0002)//(011)$_{Li}$) and <10-10>$_H$/<113>$_B$ with {0002}$_H$/{110}$_B$ described by Zhang et al. has near Burger OR. Ti and Li crystal structure following the Burger OR is shown Figure S8.

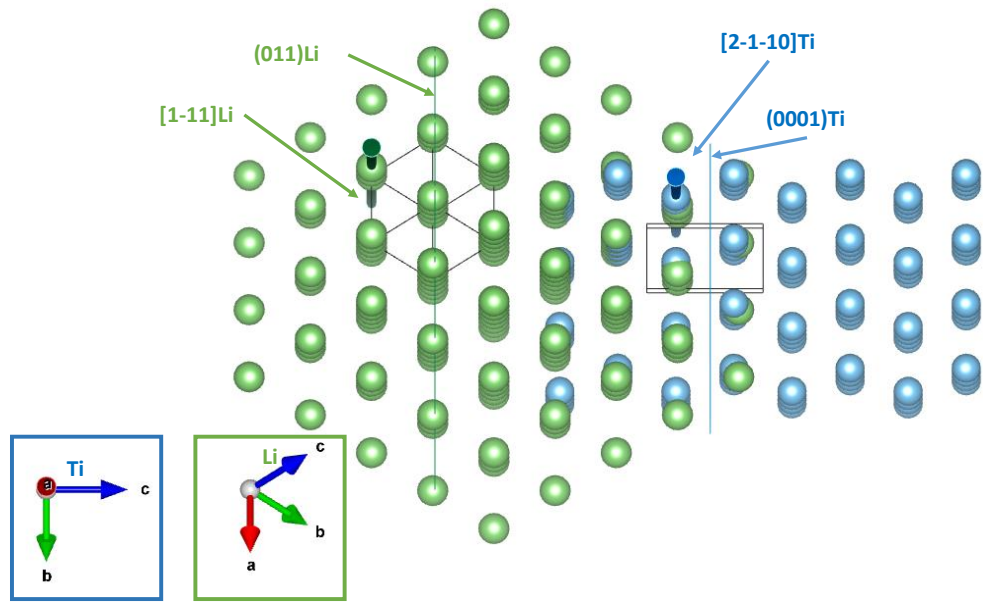

**Figure S8. Ti and Li metal atomic structure aligned according to the Burger orientation relationship. Ti and Li atoms are in blue and green respectively. (110)$_{Li}$ and (0001)$_{Ti}$ planes and [1-11]$_{Li}$ and [2-1-10]$_{Ti}$ are represented.**

How would the OR translate into XRD diffraction? Ti current collector is textured. We observed the (100) diffraction spot at azimuth 30°, 90°, and 150° and the (002) spots at 0° and 180°. The resulting preferred orientations of Ti are shown Figure S9 with respect to the film orientation. If Li would grow on Ti following the predicted Burger OR, (011)$_{Li}$ planes would be orientated as described on the bottom panel of Figure S9 hence with (011) diffraction spots at 0°, 60°, 120°, 80° and 240° which is what is observed at the beginning of plating on Figure 4.

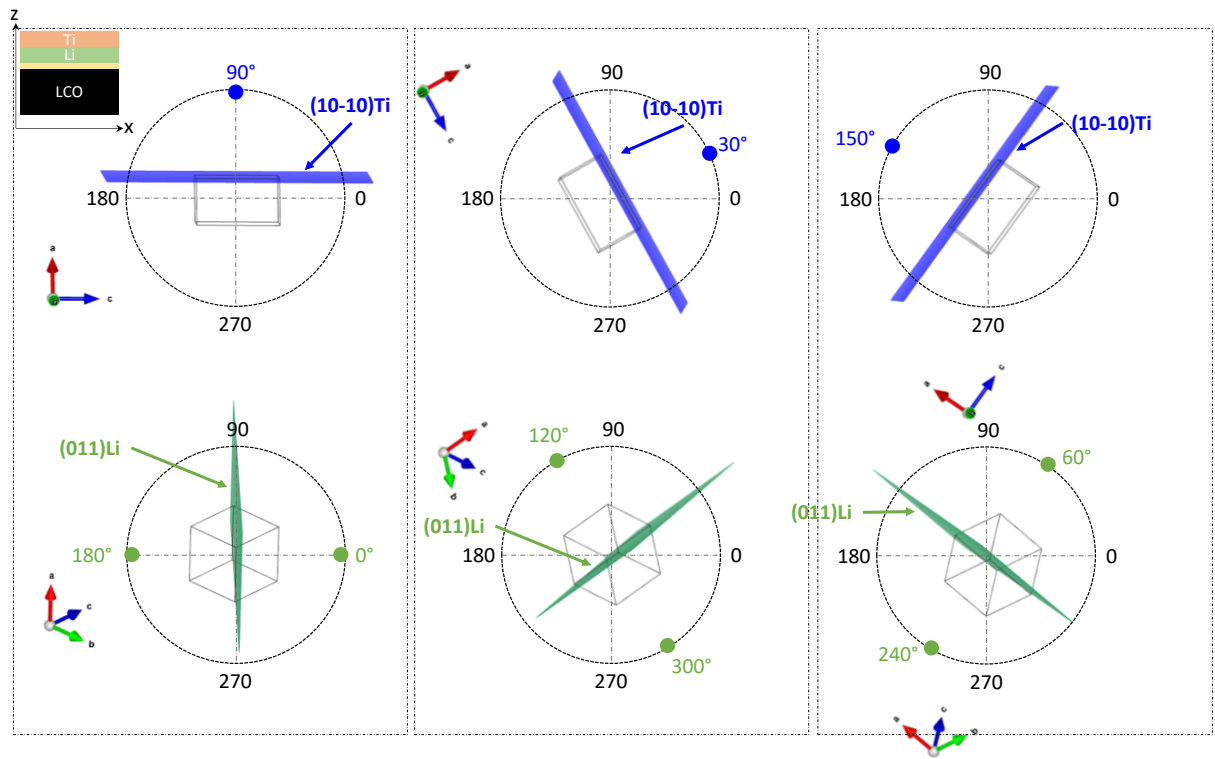

**Figure S9. Top three schematics represent different orientations of the (10-10)$_{Ti}$ and the resulting expected diffraction spot signal observed at 90°, 30° and 150° (marked as a blue dot). Bottom three schematics shows the different orientations of the (011)$_{Li}$ in Burger OR with the (10-10)$_{Ti}$ and the resulting expected spots on the detector at 0°, 30°, 60°, 180°, 240°, and 300°.**

Limitations of the current measurements with regard to the OR of Ti and Li. Using µXRD imaging without sample rotation, we probe a small range of the reciprocal space. As a consequence, we have partial information (1) see only grains in diffraction conditions, (2) see only a single diffraction spot for each grains. The second point implies that we do not completely resolve the grain orientation. Indeed, we are insensitive to the angle of the crystal around the probed scattering vector. For example, if we measure the (110) diffraction spot, we do not know the rotation angle of the plane (110) around the c axis. However, the observed pattern in the distribution of azimuth of (110)$_{Li}$ (presence of spots around 0°, 60°, 120°, 180° etc. at the beginning of plating) is clearly observed and shows the presence of a texture in Li metal. This pattern can be explained by the predicted OR of Li and Ti. A direct evidence of the presence of this OR would be possible by performing a different experiment probing a larger reciprocal space for example. However, technics capable of measuring crystal orientation, *in situ*, in a buried layer (Li grows inside the battery and is not directly accessible), with spatial resolution are rare. Some of the possible options are (1) TEM observation of an ion-milled battery, (2) Synchrotron X-ray pole figure, (3) X-ray microLaue. Note that these three technics are extremely difficult to perform because (1) It is unclear if it is possible to cut a battery without inducing morphological changes in the Li metal layer, (2) Pole figure and microLaue would have to be measured in Li through Ti current collector and Au RDL which is a challenge due to the small scattering cross section of Li compared to these two other elements. Therefore, applying these technics would be of great interest but requires challenging technical developments together with data analysis challenge.

**8 – Optical image of the battery before and after cycling**

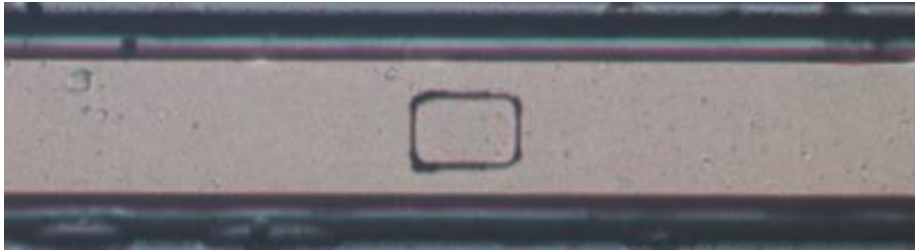

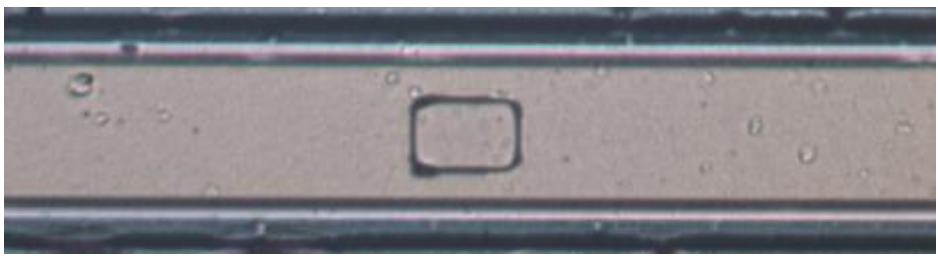

**Figure S10. Optical image of the battery after cycling. The top surface of the battery looks distorted:** out-of-fab and cycle optical image of a battery

## 9 - Discussion on the effect of external parameters (pressure, current density etc.) on texture.

Our data suggest an epitaxial growth of Li metal on Ti current collector at the beginning of charge. All literature, to be best of our knowledge, on Li metal texture is dedicated to batteries with liquid electrolytes. This is presumably due to the difficulty of probing buried Li metal texture in all solid-state systems. For Li-ion liquid batteries, *ex situ* cryoTEM and pole figures obtained from X-ray diffraction, or *operando* grazing incidence WAXS have been performed, and some consensus seem to emerge pointing that the areal capacity, nature of the SEI and of the substrate play a role on the Li texture and microstructure. Homoepitaxial growth is observed on clean Li metal surfaces[5,6] or at low stacking pressures[7]. Classical Cu current collector does not seem to induced heteroepitaxial growth[8] presumably due to the formation of native layer at the Cu surface. Also, [110] oriented Li can be grown onto [200] Li layer or Cu current collector when large amount of Li is deposited[9], in the case of highly conducting LiF-rich SEI[10] and/or using high stack pressures [7]. Generally, textured Li layer are advantageous for battery performances due to two main reasons. First, epitaxial Li deposition, whatever its texture, usually leads to dense Li layer which is better for capacity retention because less likely to form dead Li[5]. Second, [110] oriented Li surface shows a lower charge transfer resistance and higher in-plane Li diffusion leading to less dendrite formation and smaller polarisation [9]. In our battery system, we find heteroepitaxial growth of Li and Ti which means that the Li/LIPON interface doesn't prevent the orientation relationship between Li and Ti to develop. Li/LIPON interface has been reported to be composed of $Li_3P$, $Li_3N$, $Li_2O$, $Li_3PO_4$ nanoclusters embedded into an amorphous matrix [11,12]. Our device, which is anodeless, is different from previous reports, and hence a follow-up study to track the impact of the nature of the SEI in AEA anode less batteries on Li morphology and texture at the beginning of plating could be interesting. We also find that the epitaxial relationship between the Ti and Li is lost after 0.8 mAh.cm² capacity, which is lower compared to other work in Li-ion liquid batteries (> 10 mAh.cm²) despite having close current densities (around 0.4 mAh.cm²). We hypothesize that this could be due to higher stacking pressure in our all solid state battery device. Finally, despite epitaxial Li/Ti growth, the Li metal layer formed in this study is weakly textured calling for further study on highly textured Ti layers, pressure and/or current density to promote an epitaxial growth of highly [110] oriented Li layer.

## 10 - Discussion on the grain dynamics.

We unveiled dynamic information on grain size and number (1) there is a large number of grains at the beginning of charge, and the total grain number keeps decreasing during the rest of plating, striping and rest, and (2) Li grains move during plating, striping, and also during rest to a lesser extent. Motoyama et al. has performed *in situ* and *ex situ* SEM experiments looking at the deformation of the current collector when Li deposits underneath in anodeless all solid state microbatteries. This observation allows to count the nucleation points, and the authors showed that increasing the current density leads to more Li metal nucleation points and hence a large number of grains [13,14]. This trend is also observed in liquid based batteries [8]. In liquid batteries, stack pressure and capacity has been shown to increase grain size [7,9]. Indeed, Zhao et al. showed very clear *ex situ* SEM images of the Li metal grains with different areal capacities ranging from 2 mAh.cm² to 20 mAh.cm² with Li grains size growing from 1.8 μm to 6.4 μm in the in plane direction, suggesting some kind of growth mechanism for which little information is available. In light of these reports, the evolution of the number of grains in diffraction conditions can be explained. First, a high number of Li grains nucleate due to the relatively high current density at the beginning of the potentiostatic charge (1 mAh.cm²), then, as the Li metal layer grows and stack pressure increases, Li metal grains grow leading to the reduction of the grain number. Interestingly, the grain number keeps reducing at the end of charge, where the current density is quite low (0.1 mAh.cm²), or during rest at the end of discharge showing that Li metal growth mechanism is still active even without Li metal plating/striping. This is further confirmed by the observation of new grains in diffraction conditions during the rest showing the dynamic nature of the Li metal layer even at rest.

**11 – Typical detector image**

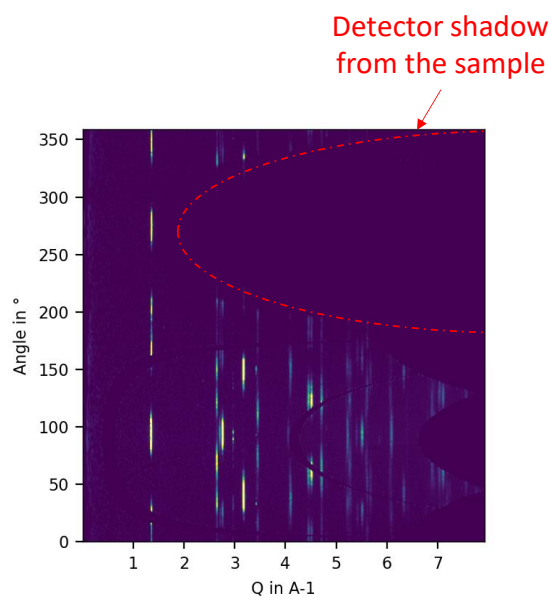

**Fig. S11. Detector image after azimuthal integration of LCO electrode showing the shadow coming from the sample at azimuth ranging between 200 – 360 °.**

**References :**

(1) Burgers, W. G. On the Process of Transition of the Cubic-Body-Centered Modification into the Hexagonal-Close-Packed Modification of Zirconium. *Physica* **1934**, *1* (7–12), 561–586. https://doi.org/10.1016/S0031-8914(34)80244-3.
(2) Potter, D. I. The Structure, Morphology and Orientation Relationship of V3N in α-Vanadium. *Journal of the Less Common Metals* **1973**, *31* (2), 299–309. https://doi.org/10.1016/0022-5088(73)90165-3.
(3) Zhang, M.-X.; Kelly, P. M. Edge-to-Edge Matching and Its Applications: Part I. Application to the Simple HCP/BCC System. *Acta Materialia* **2005**, *53* (4), 1073–1084. https://doi.org/10.1016/j.actamat.2004.11.007.
(4) Rong, W.; Dunlop, G. L. The Crystallography of Secondary Carbide Precipitation in High Speed Steel. *Acta Metallurgica* **1984**, *32* (10), 1591–1599. https://doi.org/10.1016/0001-6160(84)90218-9.

(5) Baek, M.; Kim, J.; Jeong, K.; Yang, S.; Kim, H.; Lee, J.; Kim, M.; Kim, K. J.; Choi, J. W. Naked Metallic Skin for Homo-Epitaxial Deposition in Lithium Metal Batteries. *Nat Commun* **2023**, *14* (1), 1296. https://doi.org/10.1038/s41467-023-36934-x.

(6) Xin, X.; Ito, K.; Dutta, A.; Kubo, Y. Dendrite-Free Epitaxial Growth of Lithium Metal during Charging in Li–O2 Batteries. *Angewandte Chemie International Edition* **2018**, *57* (40), 13206–13210. https://doi.org/10.1002/anie.201808154.

(7) Li, X.; Chen, C.; Fu, Z.; Wang, J.; Hu, C. Compression Promotes the Formation of {110} Textures during Homoepitaxial Deposition of Lithium. *Energy Storage Materials* **2023**, *58*, 155–164. https://doi.org/10.1016/j.ensm.2023.03.019.

(8) Wang, W.; Li, Z.; Chen, S.; Wang, Y.; He, Y.-S.; Ma, Z.-F.; Li, L. The Overlooked Role of Copper Surface Texture in Electrodeposition of Lithium Revealed by Electron Backscatter Diffraction. *ACS Energy Lett.* **2024**, *9* (1), 168–175. https://doi.org/10.1021/acsenergylett.3c02132.

(9) Zhao, Q.; Deng, Y.; Utomo, N. W.; Zheng, J.; Biswal, P.; Yin, J.; Archer, L. A. On the Crystallography and Reversibility of Lithium Electrodeposits at Ultrahigh Capacity. *Nat Commun* **2021**, *12* (1), 6034. https://doi.org/10.1038/s41467-021-26143-9.

(10) Kasse, R. M.; Geise, N. R.; Ko, J. S.; Nelson Weker, J.; Steinrück, H.-G.; Toney, M. F. Understanding Additive Controlled Lithium Morphology in Lithium Metal Batteries. *J. Mater. Chem. A* **2020**, *8* (33), 16960–16972. https://doi.org/10.1039/D0TA06020H.

(11) Cheng, D.; Wynn, T. A.; Wang, X.; Wang, S.; Zhang, M.; Shimizu, R.; Bai, S.; Nguyen, H.; Fang, C.; Kim, M.; Li, W.; Lu, B.; Kim, S. J.; Meng, Y. S. Unveiling the Stable Nature of the Solid Electrolyte Interphase between Lithium Metal and LiPON via Cryogenic Electron Microscopy. *Joule* **2020**, *4* (11), 2484–2500. https://doi.org/10.1016/j.joule.2020.08.013.

(12) Schwöbel, A.; Hausbrand, R.; Jaegermann, W. Interface Reactions between LiPON and Lithium Studied by In-Situ X-Ray Photoemission. *Solid State Ionics* **2015**, *273*, 51–54. https://doi.org/10.1016/j.ssi.2014.10.017.

(13) Motoyama, M.; Ejiri, M.; Iriyama, Y. Modeling the Nucleation and Growth of Li at Metal Current Collector/LiPON Interfaces. *J. Electrochem. Soc.* **2015**, *162* (13), A7067. https://doi.org/10.1149/2.0051513jes.

(14) Motoyama, M.; Hirota, M.; Yamamoto, T.; Iriyama, Y. Temperature Effects on Li Nucleation at Cu/LiPON Interfaces. *ACS Appl. Mater. Interfaces* **2020**, *12* (34), 38045–38053. https://doi.org/10.1021/acsami.0c02354.